\documentclass[%
reprint,
twocolumn,
superscriptaddress,
showpacs,preprintnumbers,
nofootinbib,
aps,
prd
]{revtex4-1}
\usepackage{tikz}
\usepackage{amsmath}
\usepackage{amssymb}
\usepackage{mathtools}
\usepackage{hyperref}
\usepackage{graphicx}
\usepackage{ulem}
\usepackage{comment}

\graphicspath{
{./figures/}
{./figures/boxes/}
{./figures/ratio-smallermass/}
{./figures/ratio-s1/}
{./figures/fit-xis-smallermass/}
{./figures/velocities-s1/}
{./figures/velocities-smallermass/}
}

\def\al{\alpha} 
 
\def\ga{\gamma}

\def\ep{\epsilon}

\def\th{\theta}

\def\ka{\kappa}
\def\la{\lambda}

\def\si{\sigma}
\def\ta{\tau}

\def\pa{\partial}

\newcommand{\ben}{\begin{equation}}
\newcommand{\een}{\end{equation}}
\newcommand{\bea}{\begin{eqnarray}}
\newcommand{\eea}{\end{eqnarray}}
\newcommand{\ba}{\begin{array}}
\newcommand{\ea}{\end{array}}
\newcommand{\bit}{\begin{itemize}}
\newcommand{\eit}{\end{itemize}}

\newcommand{\diag}{\mathrm{diag}\,}

\newcommand{\dBV}{d_\text{BV}}

\newcommand{\vrms}{\bar v}
\newcommand{\vrmsm}{\bar{v}_\text{m}}

\newcommand{\vrel}{\bar v_\text{rel}}
\newcommand{\xis}{\xi_\text{s}}
\newcommand{\xim}{\xi_\text{m}}
\newcommand{\xin}{\xi_\text{n}}
\newcommand{\rhos}{\rho_\text{s}}
\newcommand{\rhom}{\rho_\text{m}}

\newcommand{\mMon}{M_\text{m}}
\newcommand{\tCG}{t_\text{cg}}
\newcommand{\LatSpa}{\Delta x}
\newcommand{\tPhys}{t_{\rm p}}
\newcommand{\Swid}{w_\text{s}} % String width
\newcommand{\Mwid}{w_\text{m}} % Monopole width
\newcommand{ \tr}{\mathrm{Tr} \;}

\begin{document}

\newcommand{\Sussex}{\affiliation{
Department of Physics and Astronomy,
University of Sussex, Falmer, Brighton BN1 9QH,
U.K.}}

\newcommand{\HIPetc}{\affiliation{
Department of Physics and Helsinki Institute of Physics,
PL 64, % (Gustaf H\"{a}llstr\"{o}min katu 2),
FI-00014 University of Helsinki,
Finland
}}

\newcommand{\Stavanger}{\affiliation{
Institute of Mathematics and Natural Sciences,
University of Stavanger,
4036 Stavanger,
Norway
}}

\title{Numerical simulations of necklaces in SU(2) gauge-Higgs field theory}

\author{Mark Hindmarsh}
\email{m.b.hindmarsh@sussex.ac.uk}
\Sussex
\HIPetc
\author{Kari Rummukainen}
\email{kari.rummukainen@helsinki.fi}
\HIPetc
\author{David J. Weir}
\email{david.weir@helsinki.fi}
\HIPetc
\Stavanger

\date{April 4, 2017}

\begin{abstract}
We perform the first numerical simulations of necklaces in a
non-Abelian gauge theory.  Necklaces are composite classical solutions
which can be interpreted as monopoles trapped on strings, rather
generic structures in a Grand Unified Theory.  We generate necklaces
from random initial conditions, modelling a phase transition in the
early Universe, and study the evolution.  For all cases, we find that
the necklace system shows scaling behaviour similar to that of a
network of ordinary cosmic strings.  Furthermore, our simulations
indicate that comoving distance between the monopoles or semipoles
along the string asymptotes to a constant value at late times. This
means that while the monopole-to-string energy density ratio decreases
as the inverse of the scale factor, a horizon-size length of string
has a large number of monopoles, significantly affecting the dynamics
of string loops. We argue that gravitational wave bounds from
millisecond pulsar timing on the string tension in the Nambu-Goto
scenario are greatly relaxed.
\end{abstract}

\preprint{HIP-2016-28/TH}
\maketitle

\section{Introduction}
\label{sec:intro}

As the early universe cooled and expanded, it may have undergone
several symmetry-breaking phase transitions. Depending on the details
of the symmetry breaking, it is possible that topological defects
could have formed during such phase transitions. Probably the most
important class of topological defects for the purposes of cosmology
are cosmic strings~\cite{Kibble:1976sj} (see
Refs.~\cite{Hindmarsh:1994re,Vilenkin:2000jqa,Copeland:2011dx,Hindmarsh:2011qj}
for reviews). These are one-dimensional defects which, in the simplest
case of an Abelian Higgs model, arise from the breaking of a
$\mathrm{U}(1)$ symmetry. The resulting cosmic strings are then
extended Nielsen-Olesen vortex lines~\cite{Nielsen:1973cs}. Cosmic
strings can also arise as fundamental objects from an underlying
string
theory~\cite{Witten:1985fp,Sarangi:2002yt,Copeland:2003bj,Lizarraga:2016hpd}.

Abelian Higgs strings have been widely studied, and their
observational consequences thoroughly
explored~\cite{Laguna:1989hn,Vincent:1997cx,Moore:2001px,Bevis:2006mj,Bevis:2010gj}.
Superstrings or field theories with non-Abelian symmetries can produce
richer physics: for example, the symmetry-breaking transition
$\mathrm{SU}(2) \to Z_N$ has multiple species of strings with $N$-fold
junctions \cite{Vachaspati:1986cc}. Networks of strings with junctions
have been numerically simulated in
Refs.~\cite{Copeland:2005cy,Hindmarsh:2006qn,Urrestilla:2007yw}, and
modelled in Refs.~\cite{Leblond:2007tf,Martins:2010ma}.

In this paper we report on the first 3-dimensional numerical
simulations of a network of strings in a non-abelian gauge theory, one
with symmetry-breaking SU(2)$\to
Z_2$~\cite{Nielsen:1973cs,Hindmarsh:1985xc,deVega:1986eu,deVega:1986hm,Aryal:1987sn}.
This model is particularly attractive because it can be embedded
naturally in Grand Unified Theories (GUTs) such as SO(10)
\cite{Kibble:1982ae}, for which cosmic strings are themselves argued
to be generic~\cite{Jeannerot:2003qv}. In addition to the SO(10) case
with a single scale, our model permits two symmetry-breaking scales
with an intermediate unbroken U(1) symmetry, modelling a two-stage GUT
symmetry-breaking. In this case the first stage, SU(2)$\to$U(1),
produces 't Hooft-Polyakov
monopoles~\cite{Hooft:1974qc,Polyakov:1974ek}, and the second attaches
each monopole to two strings, both carrying half the flux.  This
combination -- of a monopole trapped on a cosmic string -- is called a
bead \cite{Hindmarsh:1985xc}, and if many such beads exist on one
string then the configuration is commonly referred to as a
necklace~\cite{Berezinsky:1997td}. For a review of these systems, see
Ref.~\cite{Kibble:2015twa}.

In \cite{Hindmarsh:2016lhy} we emphasised the importance of global
symmetries in the classification of the beads. In particular, there is
a $Z_2\times Z_2$ symmetry spontaneously broken to $Z_2$ by the string
solutions, and beads can be viewed as the resulting kinks.  We
discovered new solutions in the case where the SU(2) and U(1)
symmetry-breaking scales are degenerate, due to an enlarged discrete
global symmetry $D_4$.  Each bead splits into two ``semipoles'', and
these four semipoles can annihilate only with the corresponding
anti-semipole: in a generic configuration a semipole may not find
itself next to its antipole.

The discrete global symmetry can be further promoted to a global O(2) symmetry,
which is spontaneously broken by the string solution but not the
vacuum.  Hence semipoles dissolve and the strings carry persistent
global currents, rather like a tube of superfluid.

There is wide disagreement in the literature about how necklaces
evolve in the early universe. The necklace network is characterised by
two length scales, the average comoving monopole separation $\xim$ and
the average comoving string separation $\xis$, in terms of which the
physical energy densities $\rhom$ and $\rhos$ are
\begin{equation}
\rhom \simeq \frac{\mMon}{(a\xim)^3},\quad \rhos \simeq \frac{\mu}{(a\xis)^2},
\label{e:rhom_rhos_def}
\end{equation}
where $\mu$ is the string mass per unit length when monopoles are
absent, $\mMon$ is the monopole mass and $a$ is the scale factor.
Note that the mass of a monopole on a string is generally less than
that of a free monopole, so (\ref{e:rhom_rhos_def}) is only an
estimate of the extra energy due to the trapped monopoles.

In a normal string network, the string separation is proportional to
the horizon distance, so $\xis \propto t$, where $t$ is conformal
time: this behaviour is known as scaling.  In a scaling network, all
quantities with dimensions of length (apart from the string width)
grow in proportion to the horizon distance.
 
In a necklace, there is a new dynamically important length scale
\cite{Berezinsky:1997td}
\begin{equation}
\label{eq:dbvdefn}
\dBV = \frac{\mMon}{\mu}.
\end{equation}
The ratio of the monopole energy density to the string energy density
$r = \rho_\text{m}/\rho_\text{s}$ can be written as
\begin{equation}
\label{e:rDef}
r = \frac{\dBV}{ad},
\end{equation}
where $d = {\xim^3}/{\xis^2}$ is the average comoving separation
between monopoles along the string.  It was argued in
\cite{Berezinsky:1997td} that $r$ should grow, and it was supposed
that eventually the average monopole separation should tend to the
string width. With this assumption, $r$ would evolve quickly to a
maximum value set by the ratio of the two symmetry-breaking scales.

However, it was argued in \cite{BlancoPillado:2007zr} that this
picture underestimates the effect of monopole annihilations, which act
to reduce the number of monopoles per unit length of string $1/d$. If
monopole annihilation is efficient, their average separation along the
string should scale, so $d \propto t$, or equivalently $r \sim
\dBV/\tPhys$, where $\tPhys \propto at$ is the physical time.

Given that the total density of the necklace network is
$(1+r)\mu/(a\xis)^2$, there is a very big difference in the two
scenarios, and in particular the flux of ultra-high energy cosmic
rays, $\ga$-rays and neutrinos coming from monopole annihilation
differs by many orders of magnitude.  It is clearly important to
settle the issue.

\begin{figure*}
\begin{center}
\includegraphics[width=0.3\textwidth]{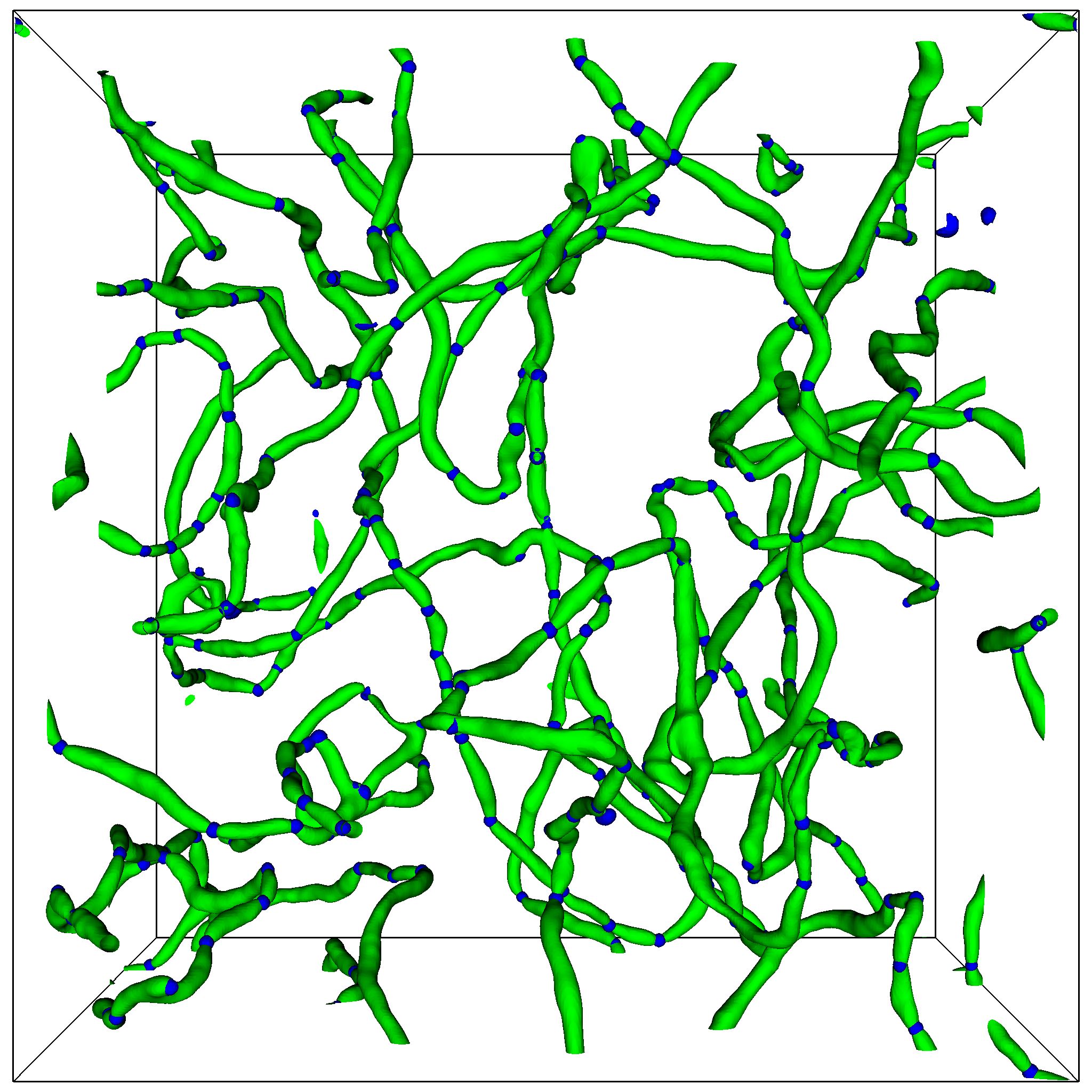}
\includegraphics[width=0.3\textwidth]{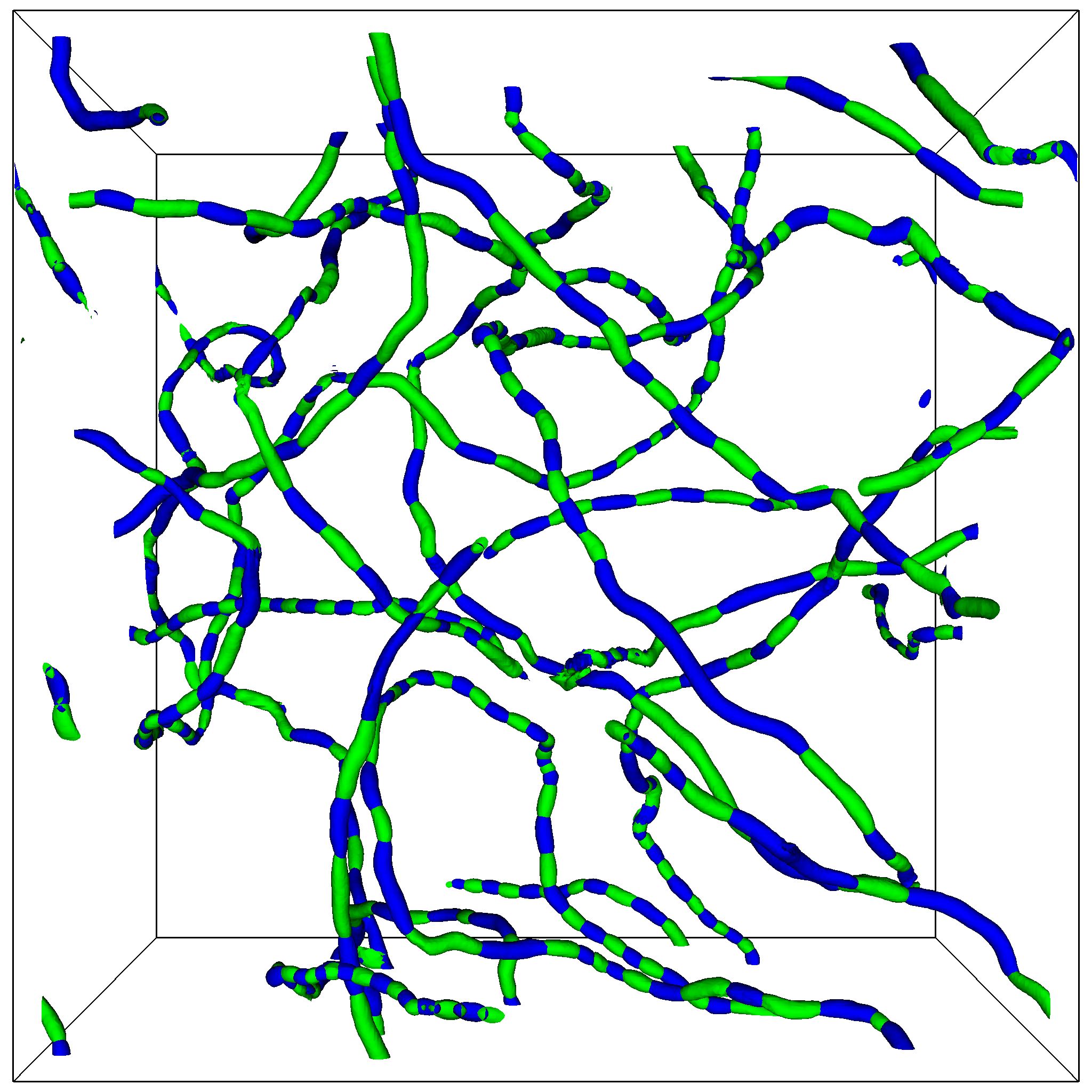}
\includegraphics[width=0.3\textwidth]{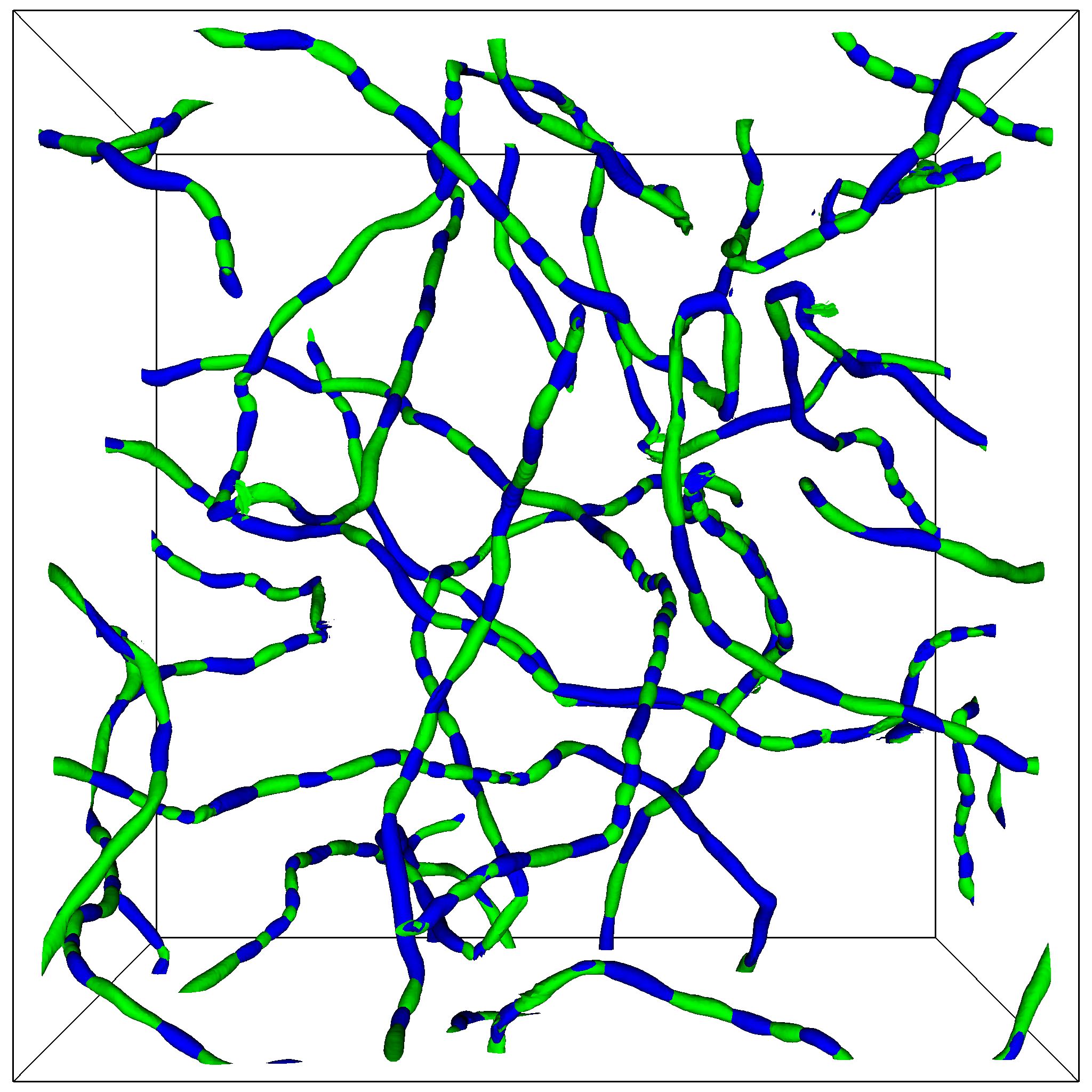}
\end{center}
\caption{\label{fig:volplot} Views of a small $360^3$ simulation for
  three different parameter choices at $t\approx 240$. The two fields
  $\Phi_1$ and $\Phi_2$ have blue and green shading respectively.  At
  left, a simulation with $m_1^2 = 0.25$ and $m_2^2 = 0.025$ (contours
  shown with $\mathrm{Tr}\,\Phi_1^2 = 0.2$ and $\mathrm{Tr}\,\Phi_2^2
  = 0.04$), giving rise to monopoles (blue) as beads on strings
  (green). The other two images show a system with $m_1^2 = m_2^2 =
  0.25$ (contours shown with $\mathrm{Tr}\,\Phi^2 = 0.2$ for both
  $\Phi_1$ and $\Phi_2$). In the centre, $\kappa=2$, showing the
  semipoles at the boundaries between the two colours. At right,
  $\kappa=1$, and the theory has a continuous global symmetry, meaning
  that the boundaries between the colours have no extra energy.}
\end{figure*}

We have performed a set of numerical simulations of a network of
strings in the SU(2)$\to Z_2$ theory (see
Fig.~\ref{fig:volplot}). They confirm the spontaneous formation of
monopoles, semipoles, and supercurrents along with the string network.

We are particularly interested in extracting the asymptotic behaviour
of the network with time, as this is essential for extrapolating to
cosmological times much later than the defect formation time.  Our
results support scaling behaviour in the total density of the necklace
network; that is, it decreases as $t^{-2}$, as does a conventional
cosmic string network.  The string and monopole average separations
$\xis$ and $\xim$ both grow with time, and the ratio of their energy
densities $r$ decreases with time.  Our simulations, while limited in
range, indicate that the mean comoving monopole separation along the
string $d$ is approximately constant, and stays the same order of
magnitude as its value when the strings form.  We also measure the
root mean square (RMS) velocities of both strings and monopoles,
finding that both relax to a values of about 0.5. If the string mass
scale is the same as the monopole mass scale, the RMS semipole
velocity is a little higher than the string RMS velocity, indicating
some relative motion.

Necklaces with a constant comoving monopole separation are a new
possibility, which has not been considered before.  In the conclusions
we briefly discuss how such a network would alter the predictions for
important observational signals.

\section{The model and its static solutions}
\label{s:ModSol}
We study the $\mathrm{SU}(2)$ Georgi-Glashow model with two Higgs
fields in a spatially flat Robertson-Walker metric.  In comoving
coordinates and conformal time, and with scale factor $a$, the action
is
\begin{align}
\mathcal{S} &=& \int d^4x \left( -\frac{1}{4}  F_{\mu\nu}^a F^{\mu\nu a} + a^2 \sum_n \mathrm{Tr} \; [D_\mu,\Phi_n][D^\mu,\Phi_n] \right. \nonumber\\
&& \left. - a^4V(\Phi_1, \Phi_2) \right)
\label{e:SU2HLag}
\end{align}
where $D_\mu = \partial_\mu + i g A_\mu$ is the covariant derivative,
$A_\mu = A_\mu^a \tau^a$, and $\tau^a = \sigma^a/2$ where $\sigma^a$
is a Pauli matrix. The Higgs fields $\Phi_n$, $n=1,2$, are in the
adjoint representation, $\Phi_n = \phi_n^a \tau^a$. Spacetime indices
have been raised with the Minkowski metric with mostly negative
signature.

The potential is
\begin{multline}
\label{e:ScaPot}
V(\Phi_1, \Phi_2) =  m_1^2 \mathrm{Tr} \; \Phi_1^2 + \lambda (\mathrm{Tr} \;
\Phi_1^2)^2 + m_2^2 \mathrm{Tr} \; \Phi_2^2   
+ \lambda (\mathrm{Tr} \; \Phi_2^2)^2  \\
+ \kappa (\mathrm{Tr} \; \Phi_1 \Phi_2)^2,
\end{multline}
with $\la$ and $\ka$ positive.
One could add a $\mathrm{Tr}\Phi_1^2 \mathrm{Tr}\Phi_2^2$ term, and have 
separate quartic couplings for the fields. However, this would not alter 
the important dynamical features of the necklace network.

The directions of the vevs are perpendicular, because of the
$(\mathrm{Tr} \; \Phi_1 \Phi_2)^2$ term in the potential.  The system
therefore undergoes two symmetry-breaking phase transitions,
$\mathrm{SU}(2) \to \mathrm{U}(1) \to Z_2$.  The vacuum expectation
values of the two adjoint scalar fields are given by $\mathrm{Tr} \;
\Phi_{1,2}^2 = \left| m_{1,2}^2 \right|/2 \lambda$, or $v_{1,2}^2 =
\left| m_{1,2}^2 \right|/ \lambda$. The scalar masses are then
$\sqrt{2} m_{1,2}$.  Without loss of generality, we can label the
scalar fields such that $\Phi_1$ has the larger vacuum expectation
value, and is responsible for the first of the symmetry-breakings.

After the first symmetry-breaking, the theory has 't Hooft-Polyakov
monopole solutions with mass \cite{Forgacs:2005vx}
\begin{equation}
\label{eq:monopolemass}
\mMon = \frac{4\pi v_1}{g}
f_\mathrm{m}\left(\frac{2\lambda}{g^2}\right); \qquad f_\mathrm{m}(1)
\approx 1.238.
\end{equation}
After the second symmetry-breaking, the theory has string solutions, with mass per unit length
\begin{equation}
\label{eq:stringtension}
\mu = \pi v_2^2 f_\mathrm{s}\left(\frac{2\lambda}{g^2}\right), 
\end{equation}
where $f_\mathrm{s}(1) = 1$.

As described in \cite{Hindmarsh:2016lhy}, in the generic case $m_1^2 >
m_2^2$ this system has a discrete global $Z_2\times Z_2$ symmetry
$\Phi_{1} \to \pm \Phi_{1}$ and $\Phi_{2} \to \pm \Phi_{2}$.  The
string solutions break it down to $Z_2$. The resulting kinks
interpolating between the two string solutions, called beads
\cite{Hindmarsh:1985xc}, can be interpreted as 't Hooft-Polyakov
monopoles with their flux confined to two tubes.  When $m_1^2 =
m_2^2$, the global symmetry is enlarged by the transformation $\Phi_1
\to \Phi_2$ to $D_4$, the square symmetry group, which is broken to
$Z_2$ by strings. The resulting kinks are labelled by a $Z_4$
topological charge. A pair of these kinks has the same charge as a
monopole on a string, hence the name semipole.

Finally, when $m_1^2 = m_2^2$ and $\ka=\la$, there is a global O(2) symmetry 
\begin{equation}
\label{e:U1Sym}
\Phi \to e^{i\al} \Phi \quad \text{and} \quad \Phi \to \Phi^*,
\end{equation}
where $\Phi = \Phi_1 + i \Phi_2$.  The phase of the complexified
adjoint scalar $\th$, defined by $\tan \th = |\Phi_2|/|\Phi_1|$,
changes smoothly along the string.  In this case the string supports
persistent supercurrents, proportional to the gradient of the phase
along the string.

In order to achieve greater dynamic range, it is common practice in
cosmic string simulations to scale the couplings and mass parameters
with factors $a^{1-s}$, where $a$ is the cosmological scale factor and
$0 \le s \le 1$.  This is done in such a way as to keep the scalar
expectation value fixed As a result, the physical string width grows
for $s < 1$, but the string tension depends only on the ratio of the
scalar self coupling to the square of the gauge coupling, and so stays
constant.  The dynamics of a string network at $s=0$ are very similar
to those at $s=1$~\cite{Daverio:2015nva}.

By contrast, the monopole mass $\mMon$ is inversely proportional to
its radius, and so $\mMon$ and the dynamical quantity $\dBV$ both grow
throughout simulations with $s < 1$.  It is therefore not clear how
the necklaces should behave in this case: the growing mass might lead
one to expect that the monopole RMS velocity should decrease, and the
monopole density increase.  We will see however that necklaces behave
similarly with $s=0$ as they do with $s=1$.

\section{Lattice implementation}
\label{s:LatImp}

\subsection{Discretisation and initial conditions}

We simulate the system by setting temporal gauge $A_0 = 0$ and then
discretising the system on a comoving 3D spatial lattice.  The
Hamiltonian of this model in the cosmological background takes the
form
\begin{multline}
\label{e:ModHam}
H(t) = \frac{1}{2g^2a^{2(s-1)}} \sum_{x,i,a} \ep_i^a(x,t)^2 + \frac{1}{2} a^2 \sum_{x; \; n,a} \; \pi_n^a(x,t)^2 \\
 + \frac{4}{g^2a^{2(s-1)}} \sum_{x; \; i<j} \left( 1- \frac{1}{2} \text{Tr}
\; U_{i j} (x,t) \right) \\ 
- a^2 \sum_{x; \; i, n} 2 \; \mathrm{Tr}\;
\Phi_n(x)U_i(x)\Phi_n(x+\hat{\imath}) U_i^\dag (x)  \\
+ a^2 \sum_{x,n} 6 \, \mathrm{Tr} \; \Phi_n^2 + a^{4} \sum_{x} V(\Phi_1, \Phi_2)
\end{multline}
where the link matrices are $U_\mu = u^0 + i \sigma^a u^a$ with
$(u^0)^2 + u^a u^a = 1$ and
\begin{equation}
\ep_i^a = -(i/2) \tr (\si^a \dot U_i U_i^\dag).
\end{equation}
With the time-varying constants, the potential becomes
\begin{multline}
V(\Phi_1, \Phi_2) = \frac{1}{a^{2(1-s)}} \left[ m_1^2 \mathrm{Tr} \; \Phi_1^2 + \lambda (\mathrm{Tr} \;
\Phi_1^2)^2 + m_2^2 \mathrm{Tr} \; \Phi_2^2  \right. \\ 
\left. + \lambda (\mathrm{Tr} \;
\Phi_2^2)^2  + \kappa (\mathrm{Tr} \; \Phi_1 \Phi_2)^2 \right].
\end{multline}

The parameter $s$ can be chosen to be smaller than its physical value
1, in order that the comoving width of the monopoles and strings
$\Mwid \sim (a^sm_1)^{-1}$, $\Swid \sim (a^sm_2)^{-1}$, does not
shrink below the lattice spacing during the simulation
\cite{Bevis:2006mj}. This extends the time range over which a
simulation can be run.

We evolve our lattice equations of motion with a standard Leapfrog
method, and the damping term is handled using the Crank-Nicolson
method.  More details of our numerical methods can be found in
Appendix~\ref{app:lattice}.

We perform simulations with both $s=1$ and $s=0$, with two different
expansion rate parameters, defined as
\begin{equation}
\label{e:ExpRatPar}
\nu = d\ln a / d\ln t.
\end{equation}
We will see that the quantities of most interest described in the next
section behave in similar ways, justifying the use of $s=0$.

Our initial conditions for $\Phi_{1,2}$ are uniformly distributed
random values in the range $[-0.5,0.5]$ for each component
$\phi_{1,2}^a$, while for the SU(2) gauge field on the lattice we
generate a random SU(2) matrix from four Gaussian random numbers
$\{u^0, u^a\}$ which we then normalise to obtain a unitary matrix of
determinant 1.

We first run for a period of time with relatively strong damping
($\sigma = 0.25$, see Appendix~\ref{app:lattice}) before switching to
standard Hubble damping at $t_{0,\text{H}}$ (see
Table~\ref{tab:s1runs}).  The momenta at the end of the damping phase
are about a thousandth the size of those arising initially from the
random initial conditions.

The procedure of seeding random fields at each site followed by a
period of over-damped evolution is standard for modelling initial
conditions for topological defects.  The important feature is that the
correlations vanish beyond a certain length scale, which is bounded
above by the causal horizon \cite{Kibble:1976sj}.  A finite
correlation length is a sufficient condition for defects to form.  In
all numerical experiments to date, the fields subsequently evolve
towards a self-similar or scaling configuration which at large
distances is independent of the initial conditions. An explicit check
of the scaling in Abelian Higgs string simulations with two different
sets of initial conditions was made in \cite{Bevis:2010gj}, although
see also \cite{Rajantie:2008bc} for a discussion of possible scaling
violation by super-horizon correlations in truly thermal initial
conditions).

We then run with $s=-1$ for a period until time $\tCG$, during which
the comoving string width grows linearly.  After $\tCG$, $s$ is set to
its physical value $s=1$.  The reason for this period of core growth
is to accelerate the preparation of the string network: the time taken
for the fields to settle to their vacua is of order $\Mwid$ and
$\Swid$, so it is helpful to arrange for them to be small while the
fields are relaxing.  The graphs of $\Mwid$ and $\Swid$ are shown in
Appendix~\ref{app:expanding} in Fig.~\ref{f:ComCorWid}.

When $\nu=0$ or when $s=0$, there is no need for the period of core
growth, and data taking can begin at $t_{0,\text{H}}$.

Our initial conditions are designed as a compromise between removing
unwanted short-distance fluctuations and allowing the strings to form
in a reasonable time.

Due to the initial cooling period, the lattice ultraviolet modes
remain strongly suppressed during the evolution of the string network.
This is justified physically, because in the early universe the local
energy density within the strings is much larger than the energy
density of the thermal background (although, in a given volume, the
total energy of the thermal background can be larger than the energy
of the string network). Thus, the thermal modes are expected to have
little influence on the string evolution.  This also helps us to avoid
the problems associated with the thermal ultraviolet modes in
real-time lattice equations of motion~\cite{Arnold:1997yb}.

\subsection{Numerical tests}

In simulations where expansion and the Hubble damping were turned off
($\nu=0$), energy conservation was better than $0.1\%$ over the period
from $t=t_{0,\mathrm{H}} = 42.5$ to $t=720$.  The root mean square
per-site relative Gauss law violation $\overline G / \overline \rho$
never exceeded $3\times 10^{-15}$ during our simulations, approaching
this value only in the initial heavy damping phase. For more details
see Appendix~\ref{app:lattice}, and in particular
Eq.~(\ref{eq:gausslaw}).

In the expanding case with $s=1$, comoving energy conservation was
obeyed to 0.1\% for simulations with $m^2_1 = 0.25$, $m^2_2 = 0.1$,
while the relative Gauss law violation $\overline G / \overline \rho$
was at most $8 \times 10^{-4}$, a value reached at the start of the
core growth phase.

We also tested whether the lattice spacing was acceptable: if the
lattice is too coarse, velocities tend to be reduced as the kinetic
energy of a defect can be converted into radiation.\footnote{For
  methods of mitigating this energy loss see
  Refs.~\cite{Moore:1996wn,Hindmarsh:2014rka}.}  For these tests we
compared $s=0$ and $s=1$ simulations at $m_1^2 = m_2^2 = 0.25$ with
those at $m_1^2 = m_2^2 = 0.1$, both with $\kappa=2$.  The string RMS
velocities at $s=1$ differed by about $1\%$ between simulations with
different masses, suggesting that any effect of the lattice spacing on
the dynamics of the strings is minor.  However, with $s=0$, monopole
and semipole RMS velocities were as much as 10 \% higher at the lower
mass, indicating that there is some lattice friction at $m_1^2 = m_2^2
= 0.25$.  Our $s=0$ runs are therefore carried out at $m_1^2 = m_2^2 =
0.1$.
  
\section{Measurements}
\label{s:Mea}

\subsection{Network length scale}

We measure the number of monopoles $N$, the string length $L$, and
study length scales derived from them.  We obtain the number of
monopoles $N$ by computing the residual unbroken $\mathrm{U}(1)$ gauge
field using projectors derived from $\Phi_1$, the Higgs field which
forms the monopoles. From this, we can compute the divergence of the
effective magnetic field and hence the magnetic charge. We give fuller
details of the U(1) projection in Appendix~\ref{app:projectors}, based
on Ref.~\cite{Davis:2000kv}.

We compute the length of string by counting the plaquettes with a
gauge-invariant ``winding'' in the U(1) subgroups formed by projection
with the scalar field $\Phi_1$, which is the heavier one in the
non-degenerate case.

The comoving string length $L$ is then defined to be the number of
plaquettes with winding.  It is possible to include a geometric
correction to $\xi_\mathrm{s}$ to account for the fact that counting
the winding number gives the Manhattan distance along the string
rather than the true string length~(see Ref.~\cite{Scherrer:1997sq}).
We choose to omit it, which should be borne in mind when comparing to
other field theory simulations~\cite{Bevis:2006mj,Daverio:2015nva}.

\subsection{Monopole density}

Several further quantities can be derived from $N$ and $L$.  First,
the average comoving string and monopole separations,
\begin{equation}
\label{e:xismDef}
\xi_\mathrm{s} = (V/L)^{1/2} \quad \text{and} \quad \xi_\mathrm{m} = (V/N)^{1/3}.
\end{equation}

We define the average comoving monopole separation along the string
\begin{equation}
d=L/N = \frac{\xim^3}{\xis^2}.
\end{equation}
and the average number of monopoles per comoving length of string
\begin{equation}
n = N/L = 1/d.
\end{equation}
The quantity $r$ defined in (\ref{e:rDef}) can be thought of as the
number of monopoles per unit physical length relative to the length
scale $\dBV$. The string and monopole separations can be combined into
one network length scale $\xin$, defined as
\begin{equation}
\label{e:xinDef}
\frac{1}{\xin^2} = \frac{1}{\xis^2}\left( 1 + r \right).
\end{equation}
The energy density of the necklace is proportional to $\xin^{-2}$, and
when $r > 1$ the majority of the energy in the network is due to the
monopoles.

Note that in the degenerate cases $m_2^2/m_1^2 = 1$ with $\ka=1$, the
points where $\Phi_1$ vanishes recorded by our monopole search
algorithm are not special: there is no local maximum in the energy
density.  However, they can be used as convenient markers of the phase
$\th$, defined after Eq.~(\ref{e:U1Sym}).

\subsection{Monopole and string velocities}

We use the positions of the strings and monopoles to compute the
string root-mean-square (RMS) velocity $\vrms$, and the monopole RMS
velocity $\vrmsm$.

Using the projection methods discussed in
Appendix~\ref{app:projectors}, we record a list of the lattice cells
that contain magnetic charge every few timesteps. We then take these
lists for two timesteps and form a distance matrix for every pair of
monopoles in the system. If the time interval $\delta t$ is much
smaller than $\xi_\mathrm{m}$, we can assume that pairing each
monopole at the later timestep with the closest one at the earlier
timestep captures the same monopole at two different times.  On the
other hand, the time interval between measurements has to be large
enough that lattice-scale discretisation ambiguities do not induce
noise~\cite{Hindmarsh:2014rka}. We will therefore compare results for
several different $\delta t$.

There are a number of standard algorithms to find the choice of
pairings in a distance matrix that minimises the total distance.  We
used a simple `greedy' algorithm that found the smallest entry in the
entire distance matrix, then removed that monopole pair, repeating
until all monopoles at the later time were paired up. This algorithm
has the advantage of being easy to code, on the other hand it scales
as the square of the number of monopoles.

The system has periodic boundary conditions, and so a `halo' region is
included from the other side of the lattice to ensure that all
possible subluminal monopole separations will be found. Once we have
determined all the pairings, we remove spurious superluminal pairings
(typically $\lesssim 1\%$ of measurements) and use the results to
determine $\vrmsm$. We considered $\delta t = 5$, $10$ and $15$ and
found convergence in the resulting curves. We used $\delta t= 15$ for
our results.  The difference from $\delta t=10$ can be considered as a
systematic uncertainty, but in practice it is comparable to or smaller
than the random error.

For the string velocities, a very similar approach was adopted, using
the positions of the plaquettes threaded by string.  As many
plaquettes can be threaded by the strings in the system, the above
pairing and distance finding algorithms were parallelised. Even so,
determining the string velocity for a few hundred thousand plaquettes
between a pair of timesteps took about five minutes on 120 processors.
For this reason, string velocities are not computed at early times,
when the number of plaquettes becomes too large.  The corresponding
monopole measurement takes about a second, and can be performed
throughout the simulations.

\section{Results}

We run over several different parameter choices for both $s=1$ and
$s=0$. 

The parameters cover both the degenerate ($m_1^2 = m_2^2$) and
non-degenerate cases, and allow us to explore the three possible
global symmetries of the string solutions, namely $\mathrm{O}(2)$,
$D_4$, and $Z_2\times Z_2$. In the degenerate case three
cross-couplings $\ka$ are considered: the special case $\kappa =
2\lambda$ having $\mathrm{O}(2)$ symmetry, and both $\kappa >
2\lambda$ and $\kappa < 2\lambda$.  For the non-degenerate case,
having $Z_2 \times Z_2$ symmetry, we explore various ratios of $m_1^2$
to $m_2^2$.

Two different expansion rate parameters $\nu = 0.5, 1$ were chosen,
where $\nu$ is defined in Eq.~(\ref{e:ExpRatPar}).  The choice $\nu =
1$ represents a radiation-dominated universe.  While $\nu = 0.5$ does
not correspond to any realistic cosmology, it is useful to explore the
impact of different expansion rates.  Simulating in a matter dominated
background ($\nu=2$) does not give enough dynamic range for reliable
results.

All runs are carried out with $m_1^2 = 0.25$ ($s=1$) and $m_1^2 = 0.1$
($s=0$).  The parameter choices are listed in Tables \ref{tab:s1runs}
and \ref{tab:runs}.  The scale factor is normalised so that $a=1$ at
the end of the simulation.

\begin{table}[t]
	\begin{center}
		\begin{tabular}{lllll|lll|lll}
			$m_1^2$ & $m_2^2$ & $g$ & $\lambda$ & $\kappa$ & $\mMon$ & $\mu$ & $\dBV$ & $\nu$ & $t_{0,\text{H}}$ & $\tCG$ \\
			\hline
			0.25 & 0.25 & 1 & 0.5 & 2 & 11 & 1.6 & 7 & 1 & 30 & 230 \\
			0.25 & 0.25 & 1 & 0.5 & 1 & 11 & 1.6 & 7 & 1 & 30 & 230 \\
			\hline
			0.25 & 0.1 & 1 & 0.5 & 1 & 11 & 0.63 & 17.5 & 0.5 & 42.5 & 242.5 \\
			0.25 & 0.1 & 1 & 0.5 & 1 & 11 & 0.63 & 17.5 & 1 & 42.5 & 242.5 \\
			\hline
			0.25 & 0.05 & 1 & 0.5 & 1 & 11 & 0.31 & 35 & 0.5 & 60 & 260 \\
			0.25 & 0.05 & 1 & 0.5 & 1 & 11 & 0.31 & 35 & 1 & 60 & 260 \\
			\hline
		\end{tabular}
	\end{center}
	\caption{\label{tab:s1runs} List of parameters for $s=1$
          (physical) runs, with dimensionful parameters given in units
          of the lattice spacing $a$.  Potential parameters
          (\ref{e:ScaPot}) are shown along with the isolated monopole
          mass $\mMon$ and the isolated string tension $\mu$ computed
          using
          Eqs.~(\ref{eq:monopolemass})~and~(\ref{eq:stringtension}).
          The length scale $\dBV$ as computed using
          Eq.~(\ref{eq:dbvdefn}) is also shown.  Finally, we quote the
          expansion rate parameter $\nu = d\ln a/d \ln t$, the time at
          which we change to Hubble damping during our simulations,
          $t_{0,\text{H}}$, and the time at which core growth ends and
          strings and monopoles reach their true physical width
          $\tCG$.  All these simulations have lattice size 720 and
          duration 720. }
\end{table}

\begin{table}[t]
	\begin{center}
		\begin{tabular}{lllll|lll|l}
			$m_1^2$ & $m_2^2$ & $g$ & $\lambda$ & $\kappa$ & $\mMon$ & $\mu$ & $\dBV$ & $t_{0,\text{H}}$ \\
			\hline
			0.1 & 0.1 & 1 & 0.5 & 2 & 6.96 & 0.628 & 11.1 & 30 \\
			0.1 & 0.1 & 1 & 0.5 & 1 & 6.96 & 0.628 & 11.1 & 30 \\
			0.1 & 0.1 & 1 & 0.5 & 0.5 & 6.96 & 0.628 & 11.1 & 30 \\
			\hline
			0.1 & 0.04 & 1 & 0.5 & 1 & 6.96 & 0.251 & 27.7 & 67.1 \\
			0.1 & 0.02 & 1 & 0.5 & 1 & 6.96 & 0.126 & 55.4 & 94.9 \\
			0.1 & 0.01 & 1 & 0.5 & 1 & 6.96 & 0.0628 & 111 & 134 \\
			\hline
		\end{tabular}
	\end{center}
	\caption{\label{tab:runs} List of simulation parameters for
          runs with $s=0$, as for Table \ref{tab:s1runs}.  The
          expansion rate parameter is $\nu=1$ (radiation era) for all
          simulations.  At $s=0$ the physical size of the monopole and
          string cores grows in proportion to the scale factor.  All
          these simulations have lattice size 720 and duration 720.}
\end{table}

The units are defined such that the lattice spacing $\LatSpa$ is 1.
All simulations are carried out on a $720^3$ lattice, with timestep
$\Delta t = 0.25$ after the initial heavy damping period ends at
$t_{0,\text{H}}$, for a total time $720$, or one light-crossing time
of the box.  In principle, correlations can start to be established
after half a light-crossing time. However, the only massless
excitations are waves on the string, and the strings are much longer
than the box size even at the end of the simulations.  The network
length scale does not show any evidence for finite-size effects,
although it is possible that the slight increase in $d$ for semipoles
and supercurrents at $t \gtrsim 360$ in Fig.~\ref{f:n_both} is a sign
of the limited simulation volume.

Each set of parameter choices is run for 3 different realisations of
the initial conditions, and our results are statistical averages over
these runs.

We investigate the monopole density with the two different measures
introduced in Section \ref{s:Mea}, the monopole-to-string density
ratio $r$ and the number of monopoles per unit comoving length of
string $n$.

\subsection{Network length scale}

\begin{figure}
\begin{center}
\includegraphics[clip=true,width=0.5\textwidth]{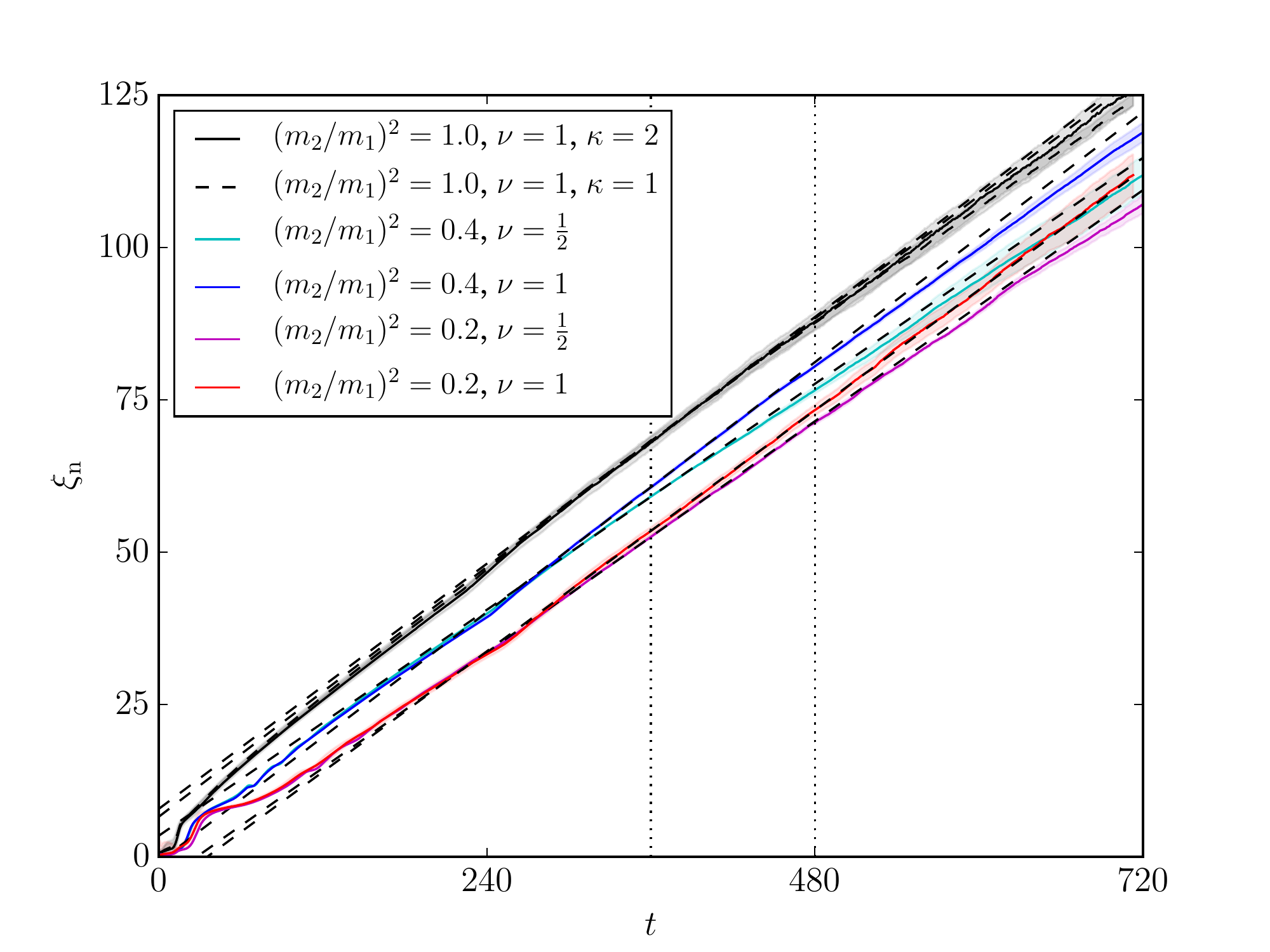}
\includegraphics[clip=true,width=0.5\textwidth]{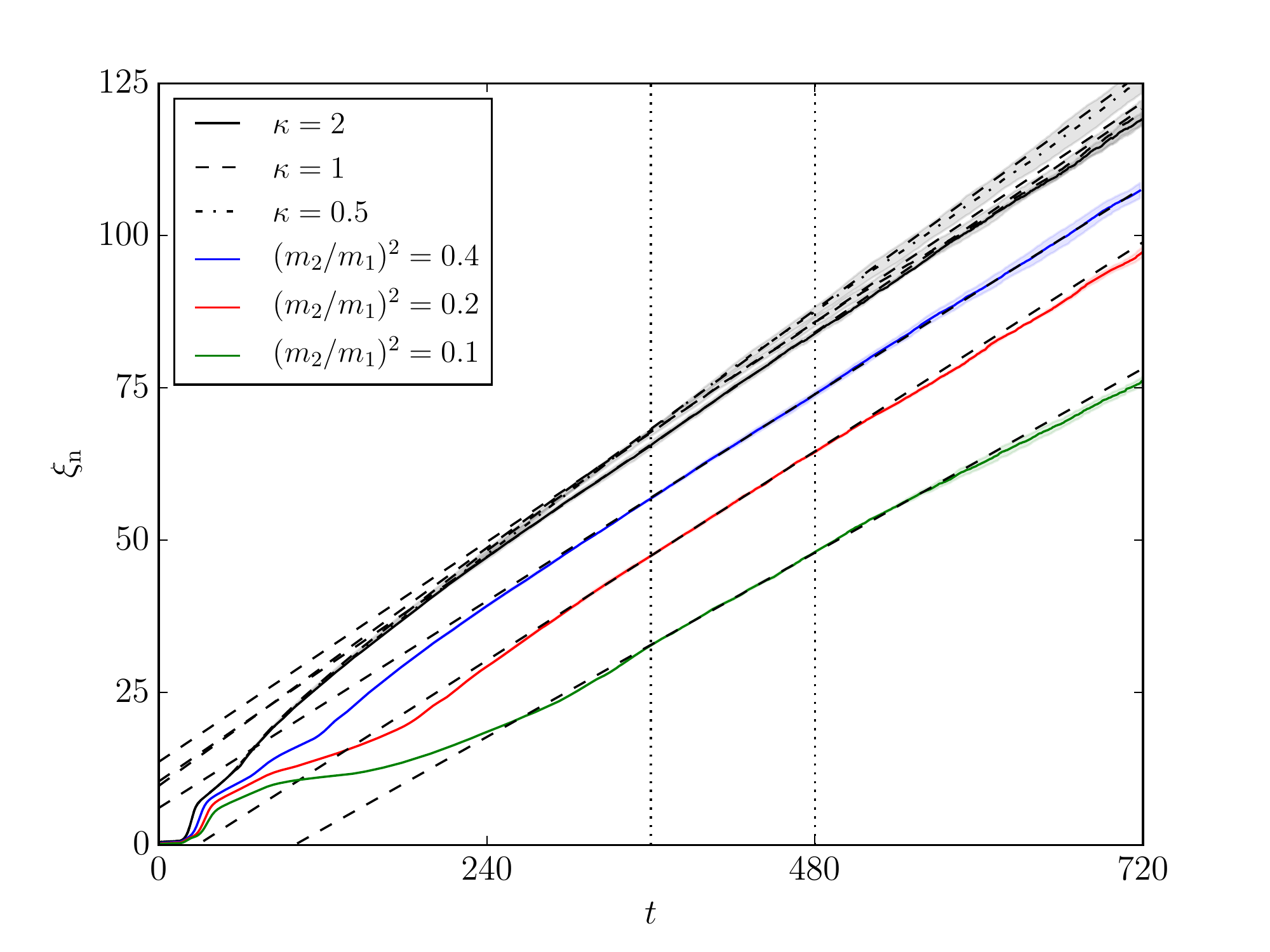}
\end{center}
\caption{\label{fig:xin} Plot of the network length scale $\xin$,
  defined in Eq.~(\ref{e:xinDef}), with core growth parameter $s=1$
  (top) and $s=0$ (bottom).  Fits to linear growth are also shown,
  within the range indicated by the vertical dashed lines.  The
  gradients of the fit are given in Tables \ref{tab:fits_s1} and
  \ref{tab:fits_s0}.}
\end{figure}

In Fig.~\ref{fig:xin} we plot the comoving necklace network length
scale $\xin$, defined in Eq.~(\ref{e:xinDef}), for $s=1$ (top) and
$s=0$ (bottom).

All cases show linear growth with time, which means that the network
is scaling.  We perform fits in the range $360 < t < 480$, which while
in excess of the half light crossing time for the system, allows time
for the scaling behaviour to develop.  There are small differences in
the slope between simulations with different mass ratios, although
there is not enough dynamic range to ensure that they are not
inherited from differences in the initial conditions.  There is also
evidence that the lower expansion rate $\nu = 1/2$ the slope is lower,
i.e. that the average necklace density is higher.

\begin{table}[h!]
\begin{center}

\begin{tabular}{lll|l|l}
$m_1^2$ & $m_2^2$ & $\kappa$ & $\nu$  & $\xi_\mathrm{n}$ gradient \\
\hline
0.25 & 0.25     & 2   & 1 & $0.171 \pm 0.002$ \\
0.25 & 0.25     & 1  & 1 & $0.168 \pm 0.004$ \\
\hline
0.25 & 0.1   & 1 & 0.5 & $0.154 \pm 0.001$ \\
0.25 & 0.1   & 1 & 1 & $0.171 \pm 0.002$ \\
\hline
0.25 & 0.05    & 1  & 0.5 & $0.158 \pm 0.002$ \\
0.25 & 0.05    & 1  & 1 & $0.165 \pm 0.004$ \\
\hline
\end{tabular} 
  
\end{center}
\caption{\label{tab:fits_s1} Gradients for the network comoving length
  scale $\xi_\mathrm{n}$, from the fits shown in the graphs of $\xi_n$
  against conformal time $t$ for $s=1$ in Fig.~\ref{fig:xin} (top). }

\bigskip

\begin{tabular}{lll|l}
$m_1^2$ & $m_2^2$ & $\kappa$  & $\xi_\mathrm{n}$ gradient \\
\hline
0.1 & 0.1     & 2   & $0.154 \pm 0.005$ \\
0.1 & 0.1     & 1  & $0.150 \pm 0.003$ \\
0.1 & 0.1     & 0.5  & $0.163 \pm 0.008$ \\
\hline
0.1 & 0.04   & 1 & $0.141 \pm 0.004$ \\
0.1 & 0.02    & 1  & $0.143 \pm 0.001$ \\
0.1 & 0.01     & 1  & $0.126 \pm 0.001$ \\
\hline
\end{tabular}
\caption{\label{tab:fits_s0} Gradients for the network comoving length
  scale $\xi_\mathrm{n}$, from the fits shown in the graphs of $\xi_n$
  against conformal time $t$ for $s=0$ in Fig.~\ref{fig:xin}
  (bottom). }

\end{table}

\subsection{Monopole density}

\begin{figure}
  \begin{center}
    \includegraphics[clip=true,width=0.5\textwidth]{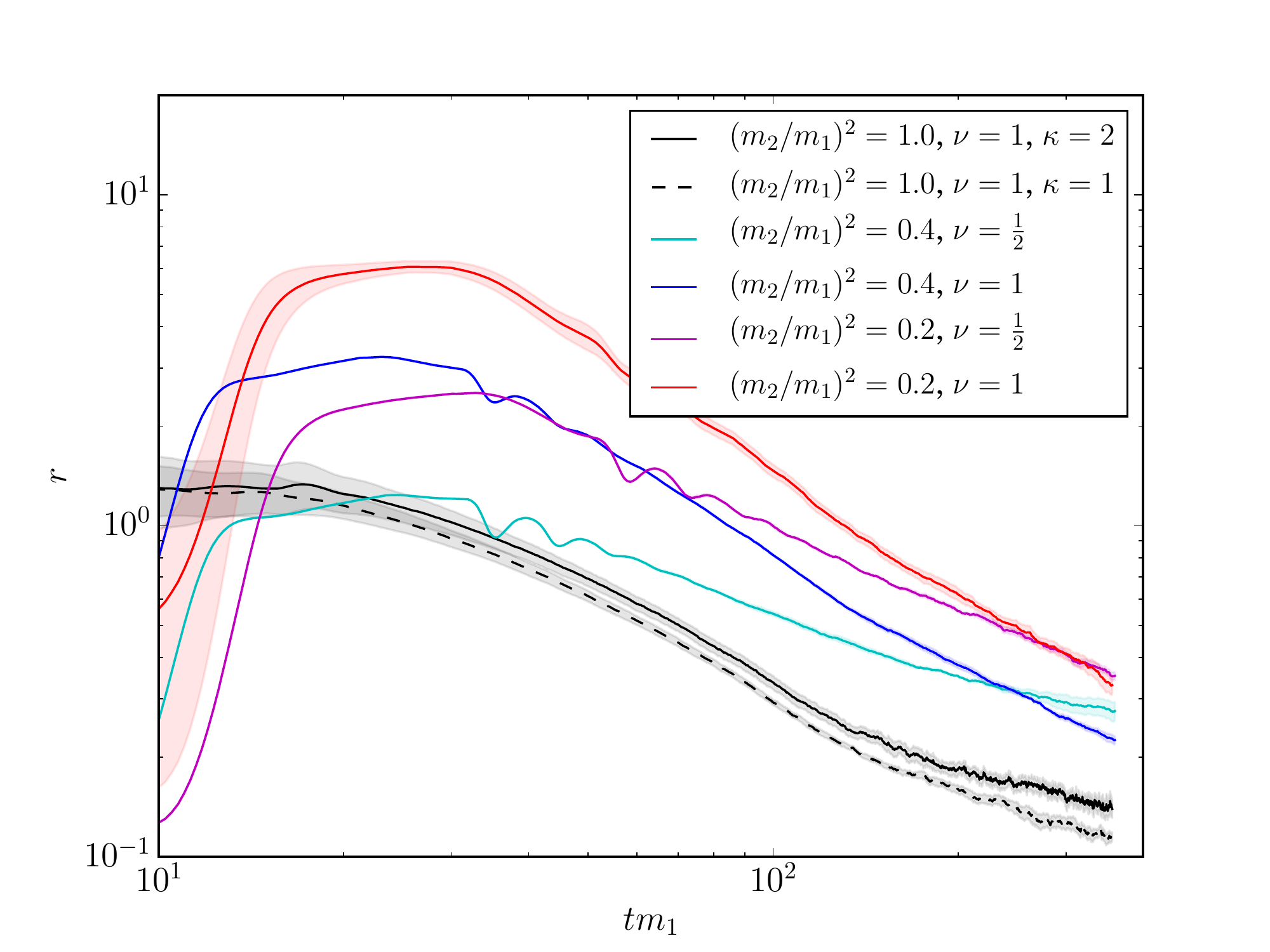}
  \end{center}
  \caption{\label{f:r_s1} The ratio of monopole to string energy
    density (\ref{e:rDef}) in simulations with $s=1$.  The legend
    gives the expansion rate parameter $\nu = d \log a/d \log t$, the
    mass ratio of the fields $m_2/m_1$, and in the degenerate case the
    value of the cross-coupling $\ka$, which is otherwise $\ka=1$.
    The mass parameter $m_1^2 = 0.25$. }
\end{figure}

In Fig.~\ref{f:r_s1} we plot the ratio of monopole to string energy
density $r$, defined in (\ref{e:rDef}), against time in units of
$m_1^{-1}$, for all parameters given in Table \ref{tab:s1runs}.  Note
that $m_1^{-1}$ is approximately the monopole size.

We see that $r$ decreases after the formation of the string network,
with what appears to be a power law after the core growth period has
finished.

\begin{figure}
\begin{center}
\includegraphics[clip=true,width=0.5\textwidth]{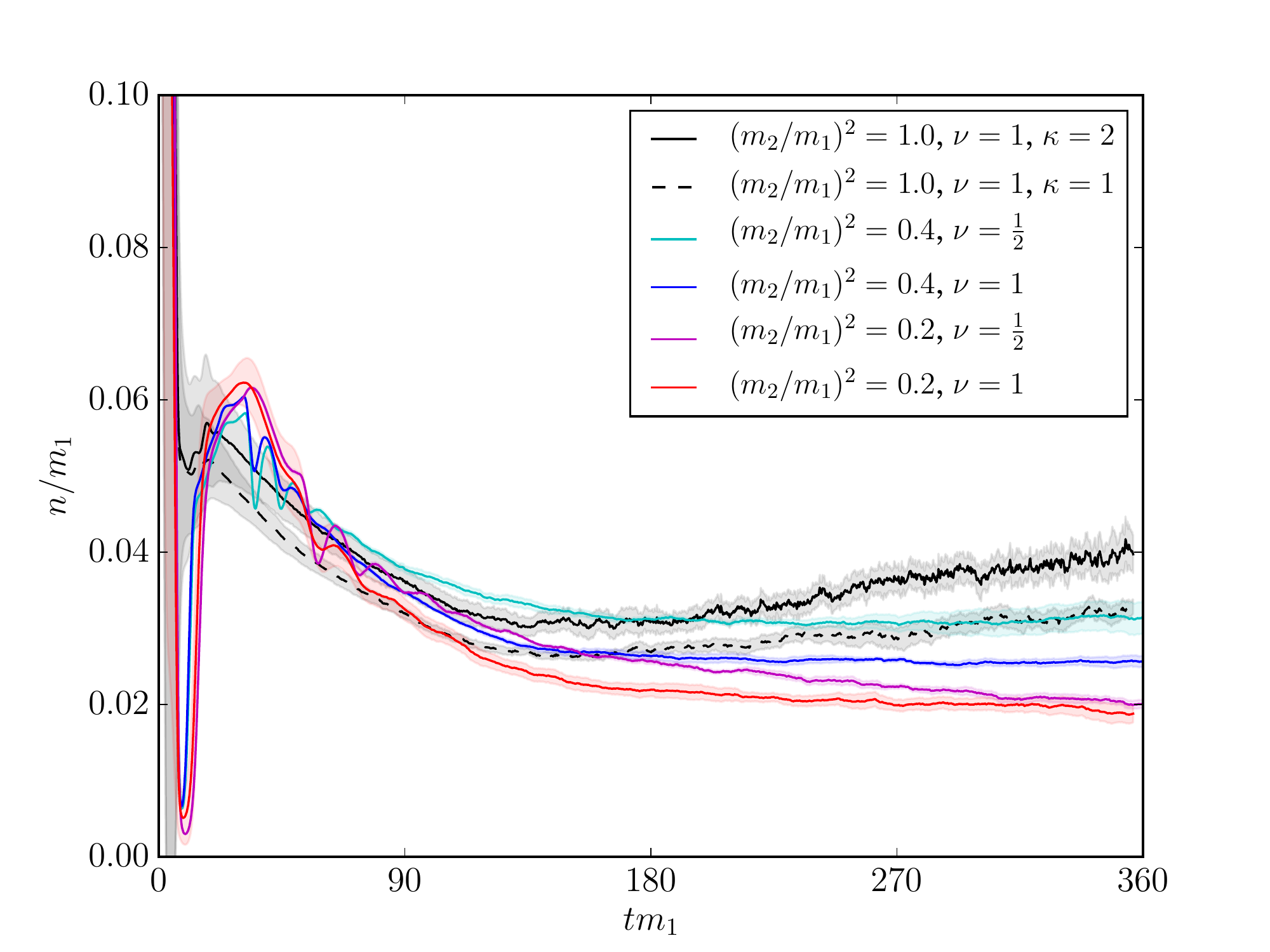}
\includegraphics[clip=true,width=0.5\textwidth]{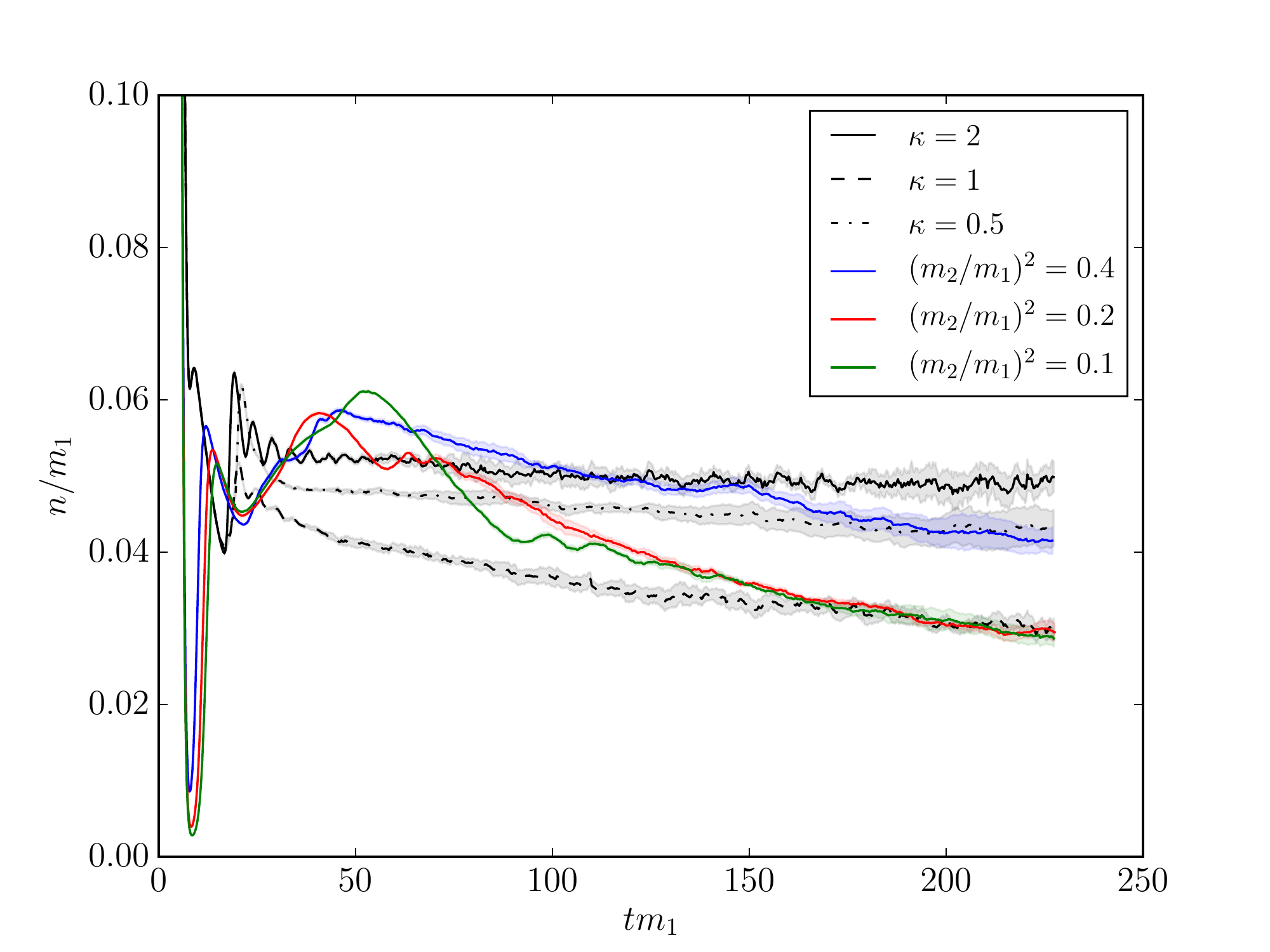}
\end{center}
\caption{\label{f:n_both} The number of monopoles per comoving string
  length in simulations with $s=1$ (top) and $s=0$ (bottom). The
  legend gives the expansion rate parameter $\nu = d \log a/d \log t$,
  the mass ratio of the fields $m_2/m_1$, and in the degenerate case
  the value of the cross-coupling $\ka$, which is otherwise $\ka=1$.
  The mass parameter $m_1^2 = 0.25$ ($s=1$) and $m_1^2 = 0.1$
  ($s=0$).}
\end{figure}

The significance of the power law is clearer if we plot the comoving
linear monopole density on the string $n$, again in units of
$m_1^{-1}$ (Fig.~\ref{f:n_both}).  We can see from the figure that,
with the possible exception of the mass-degenerate cases ($m_2^2/m_1^2
= 1$) at $s=1$, $n$ appears to tend to a constant at large time.
Hence the comoving separation of the monopoles remains the same order
of magnitude as its value at the formation of the strings.

There is some evidence for a slow increase in $n$ for the degenerate
cases $m_2^2/m_1^2 = 1$ at $s=1$, which may be due to semipole
annihilations being less probable than monopole-antimonopole
annihilations -- some pairings of semipoles cannot
annihilate~\cite{Hindmarsh:2016lhy}.  However, the increase occurs
after a half-light crossing time for the simulation box, so this may
be a finite volume effect.

\begin{figure}
  \begin{center}
    \includegraphics[width=0.3\textwidth]{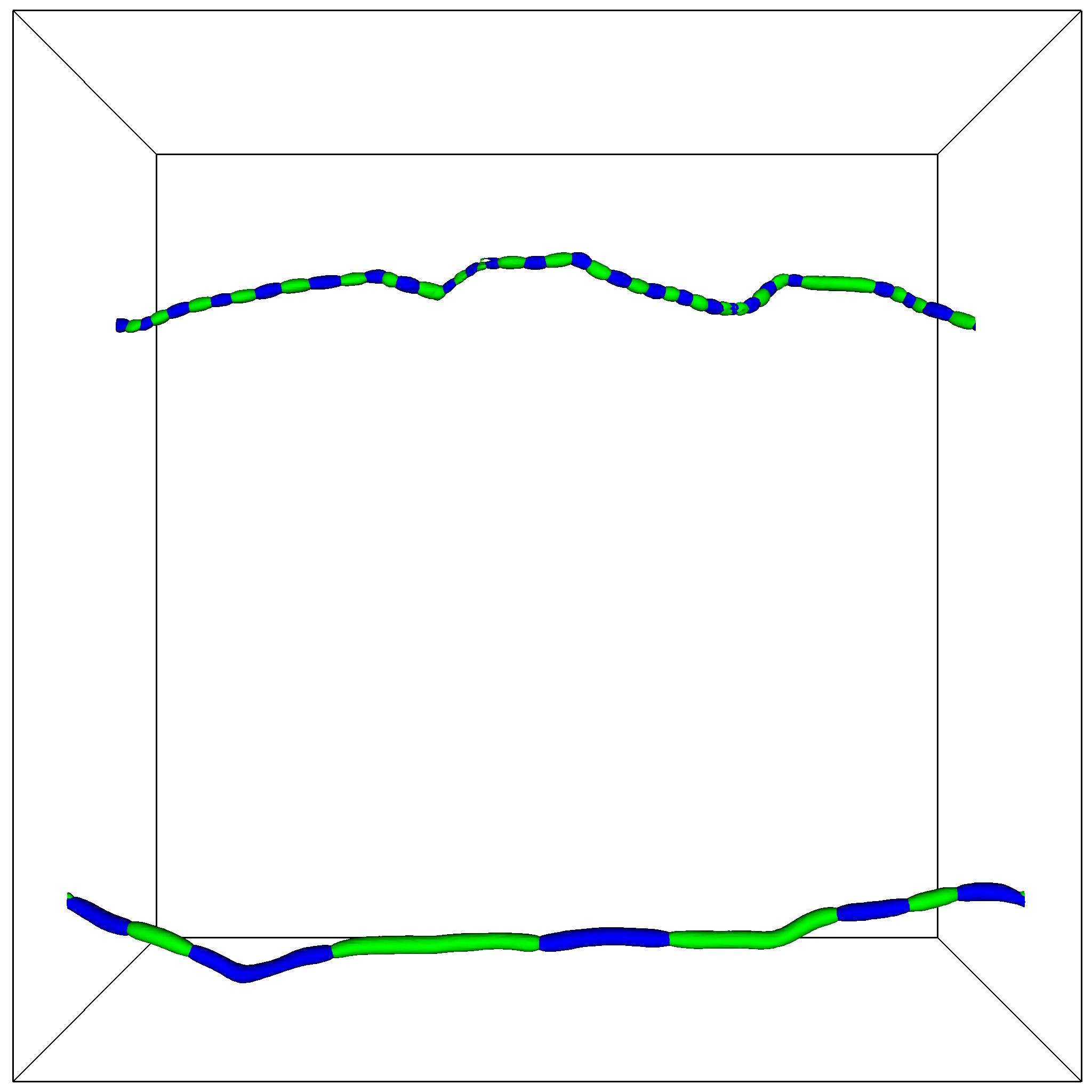}
  \end{center}
\caption{\label{f:semipole_end} A small $360^3$ box at $t = 1080$,
  simulated at $m_2^2/m_1^2 = 1$ and $\kappa = 2$.  The high density
  of semipoles on one of the strings shows that semipoles can avoid
  annihilation in some cases.  }
\end{figure}

We illustrate the ability of semipoles to avoid annihilation in
Fig.~\ref{f:semipole_end}, which depicts two strings winding around
the periodic lattice when the total length of string and the semipole
number has stabilised.  One can see that on one of the strings, the
semipole density is much higher, and examination of multiple snapshots
prior to this one shows that semipoles have repelled each other.
However, the high semipole density may be an artefact of the periodic
boundary conditions, which have prevented the strings from shrinking
in length any further.  Without this shrinking, semipoles are not
forced together, so there is less likelihood of overcoming the
repulsion and annihilating.

In the degenerate cases $m_2^2/m_1^2 = 1$ with $\ka=1$, we recall that
the recorded monopole positions are just places where the phase of the
complexified scalar has the value $\th=\pm\pi/2$.  The fact that the
comoving distance between these points remains approximately constant
indicates that the comoving RMS current is constant, and so the
physical RMS current decreases in inverse proportion to the scale
factor.

\begin{figure}
\begin{center}
\includegraphics[clip=true,width=0.5\textwidth]{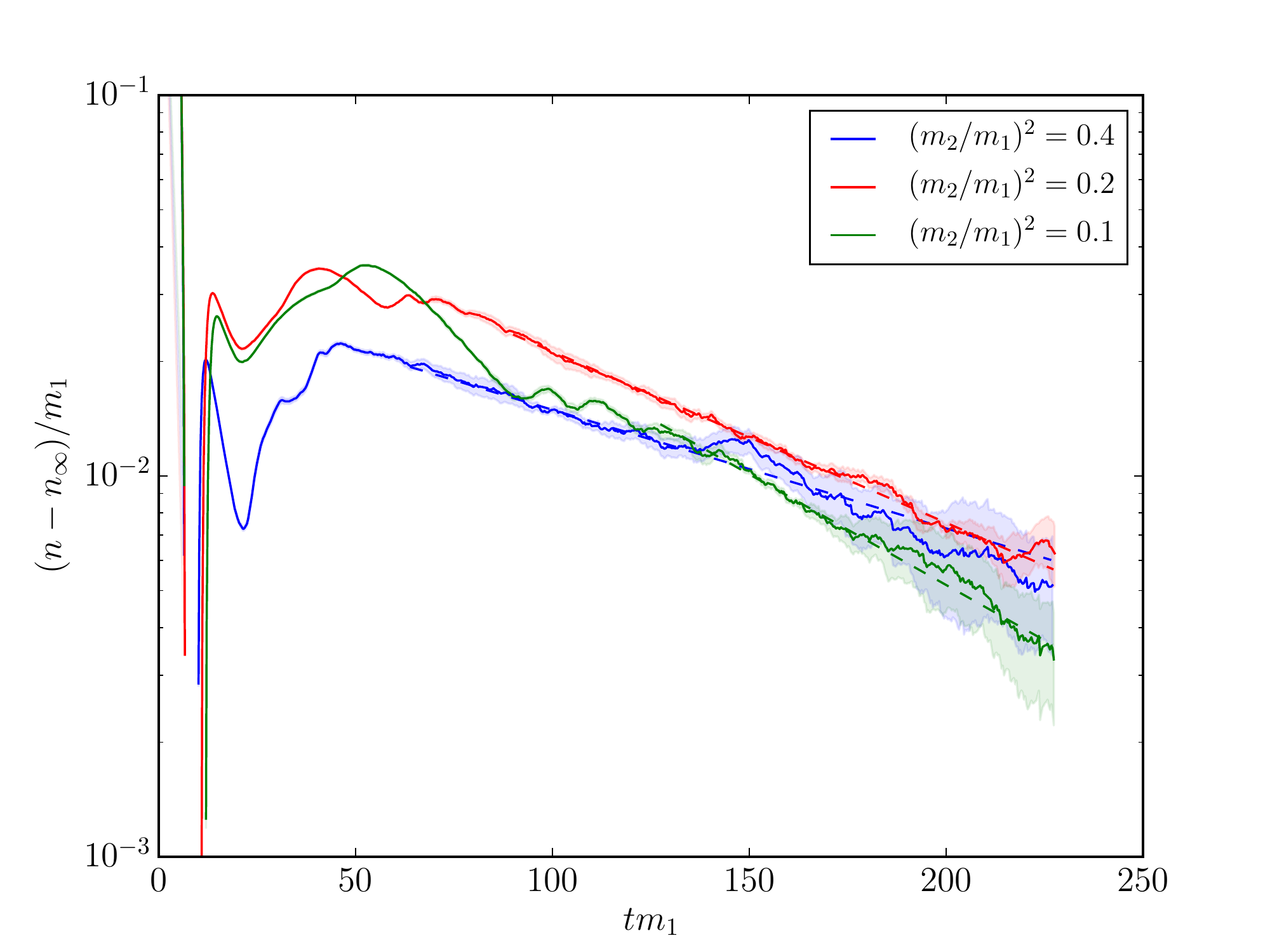}
\end{center}
\caption{\label{fig:nfit} The difference of the linear monopole
  density $n$ from its asymptotic value $n_\infty$.  The parameter
  $n_\infty$ is extracted from a fit of $n$ to a constant to
  exponential decay [see Eq.~(\ref{e:nFit})]; the fits are shown as
  dashed lines.  Both $n$ and the time are scaled by $m_1$ to make
  dimensionless quantities.  Only those values of $m_2/m_1$ where a
  reliable fit is possible are shown; for other values, the change in
  $n$ is too small.  }
\end{figure}

In the $s=0$ case, the increased dynamic range means we can attempt a
meaningful fit to investigate the relaxation to the constant $n$
evolution.  In Fig.~\ref{fig:nfit}, we show a graph of $n - n_\infty$,
where the asymptotic value of the linear monopole density $n_\infty$
is taken from a fit to the functional form
\begin{equation}
\label{e:nFit}
n = n_\infty + A\exp(-B m_1 t).
\end{equation}
Fits are shown with dashed lines, and fit parameters are given in
Table \ref{t:nFitPar}.

\begin{table}[t]
\begin{center}
\begin{tabular}{ll|lll}
$m_1^2$ & $m_2^2$ & $\frac{n_\infty}{m_1}$ & $A$ & $B$  \\
\hline
0.1 & 0.04 & 0.036 & 0.031 & 0.0072 \\
0.1 & 0.02 & 0.023 & 0.060 & 0.0104 \\
0.1 & 0.01 & 0.025 & 0.075 & 0.0134 \\
\hline
\end{tabular}
\end{center}
\caption{\label{t:nFitPar} Parameters for the fit of the linear
  monopole density data in Fig.~\ref{fig:nfit} to the function
  (\ref{e:nFit}).  All simulations are radiation era, with $s=0$.  }
\end{table}

The fits confirm the visual impression that the linear monopole
density is asymptoting to a constant non-zero value, and also support
the exponential ansatz for the relaxation.

\subsection{Monopole velocities}

\begin{figure*}[t!]
	\begin{center}
		\includegraphics[clip=true,width=0.45\textwidth]{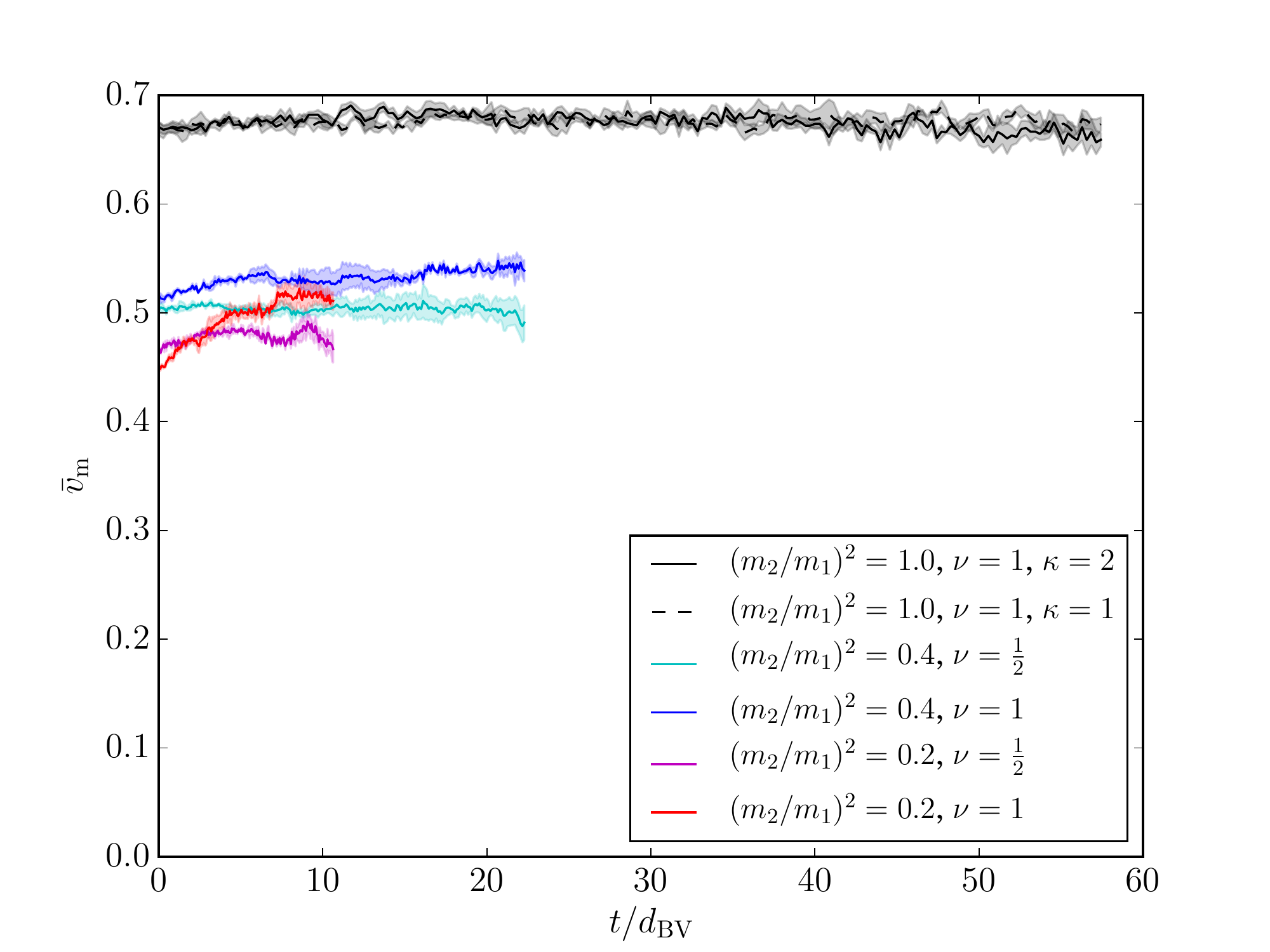}
		\includegraphics[clip=true,width=0.45\textwidth]{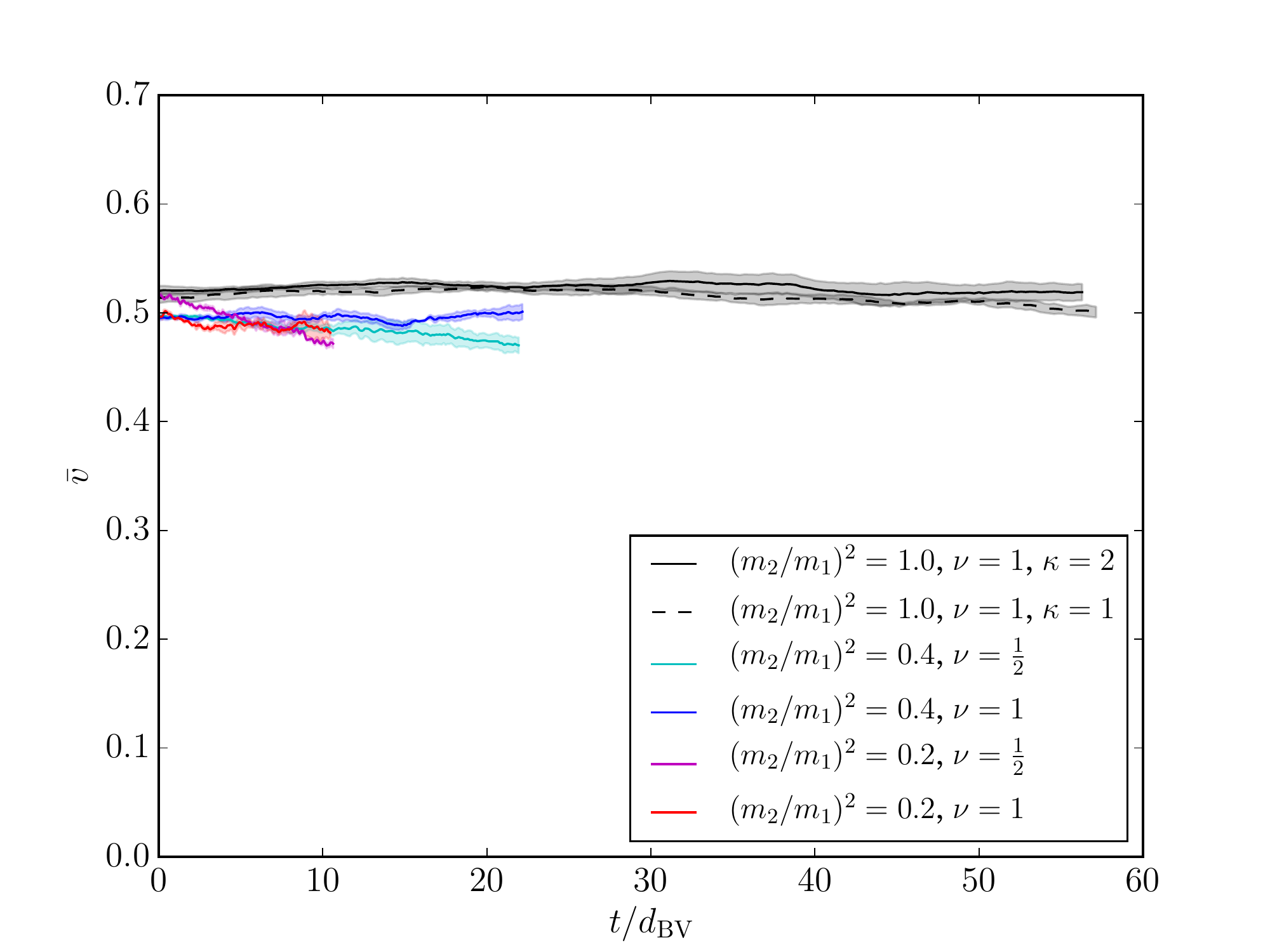}
		\includegraphics[clip=true,width=0.45\textwidth]{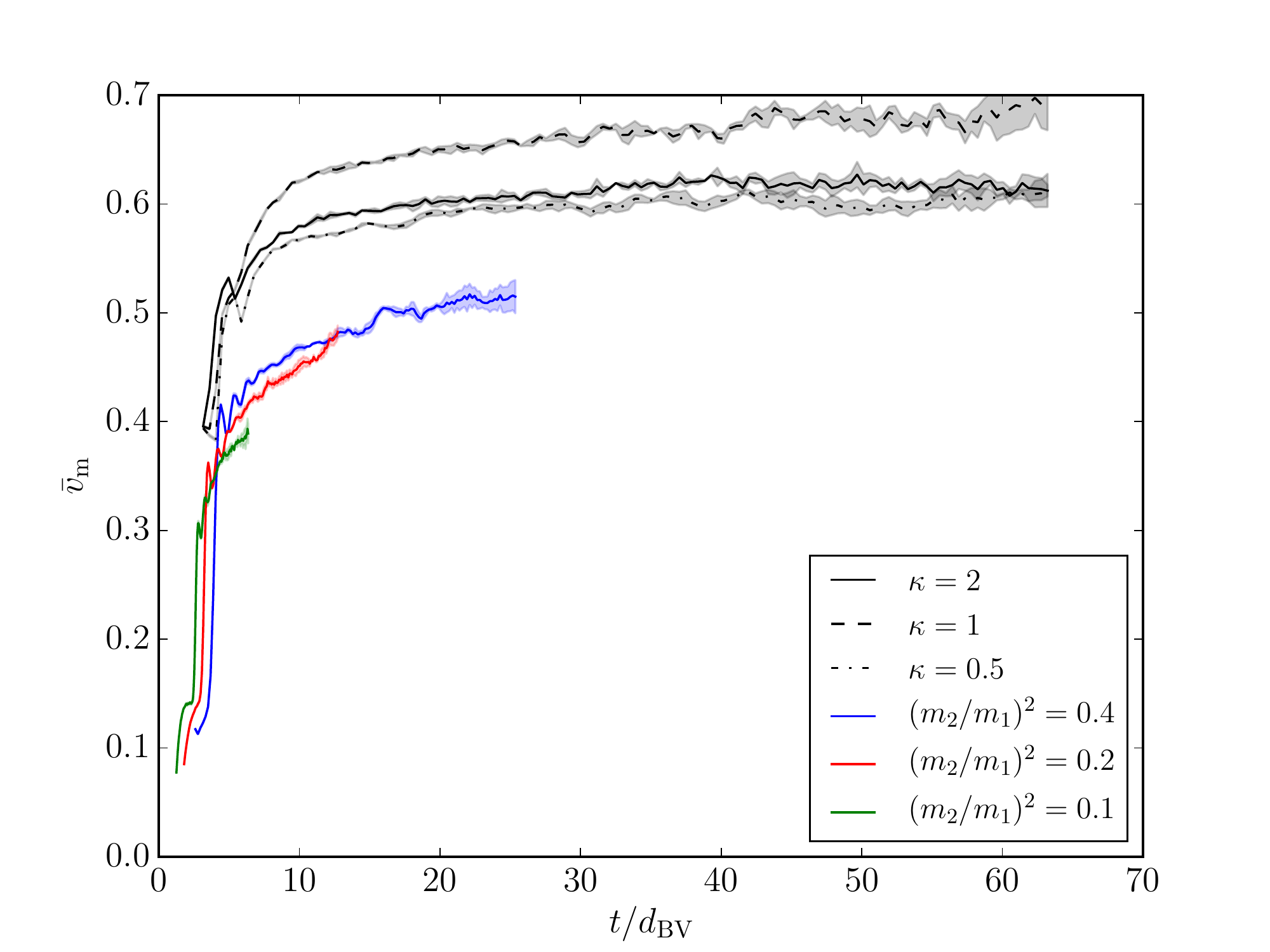}
		\includegraphics[clip=true,width=0.45\textwidth]{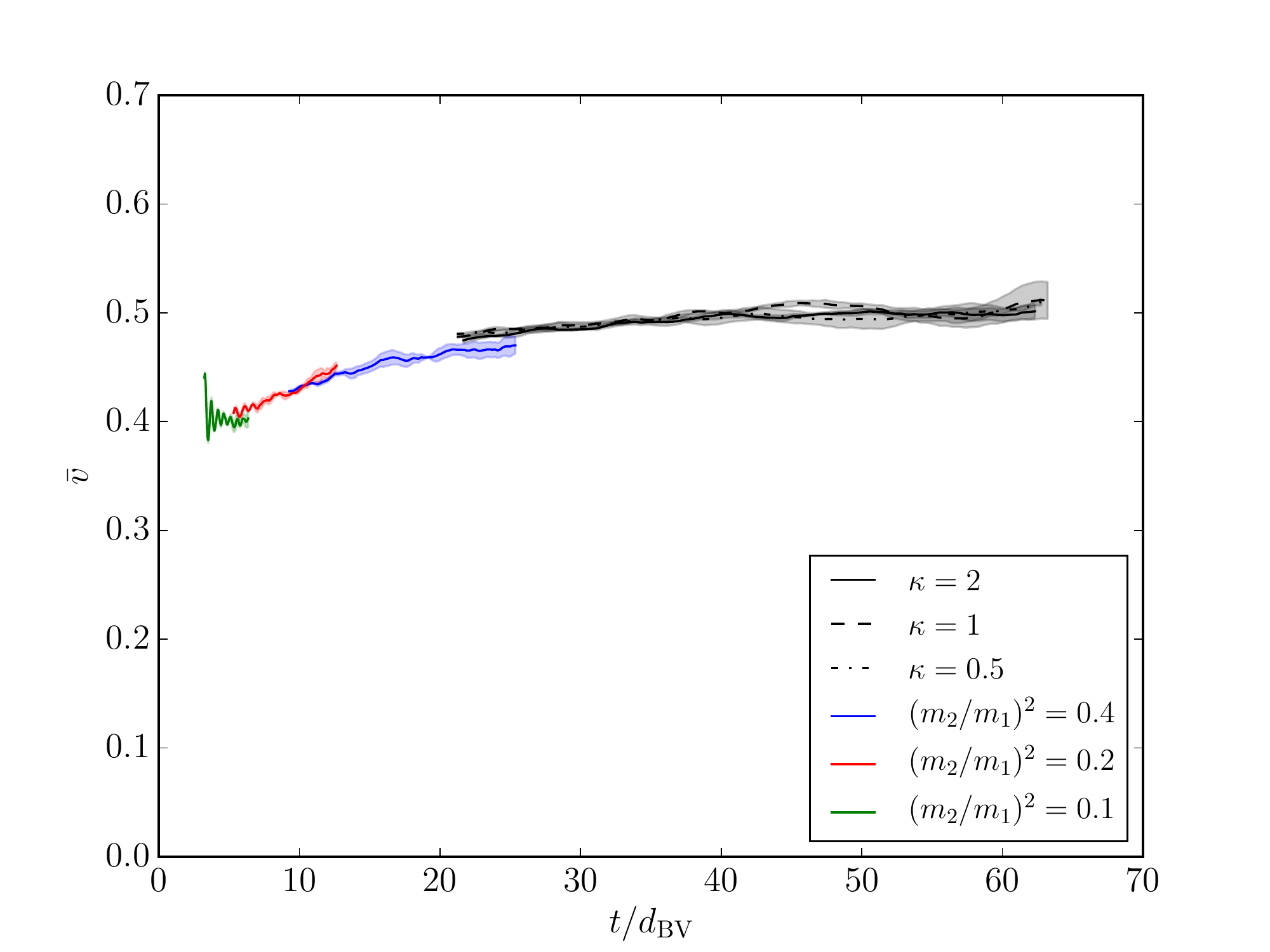}
	\end{center}
	\caption{\label{fig:rmsvels1} Plot of $\vrms$ and $\vrmsm$,
          the root mean square string and monopole/semipole
          velocities, for $s=1$ (top) and $s=0$ (bottom). The time
          axis is scaled by $\dBV$, defined in Eq.~(\ref{eq:dbvdefn}).
        }
\end{figure*}

Fig.~\ref{fig:rmsvels1} shows the RMS velocities of the strings,
monopoles and semipoles for different masses, cross-couplings $\ka$,
and expansion rate parameters $\nu$.  The RMS velocities all appear to
asymptote at the same rate $\dBV^{-1}$ to a constant value.

We see that the RMS string velocities are all around $0.5$.  When the
field mass parameters $m_1$ and $m_2$ are different, the RMS monopole
velocities are also all about 0.5, independent of the mass ratio and
the expansion rate.  If the mass parameters are the same, the RMS
monopole velocity at about $0.63$ is a little higher than the RMS
string velocity.  RMS velocities are consistent between $s=1$ and
$s=0$, with the exception of the semipoles at $s=0$, which appear to
move a little slower ($\vrmsm \simeq 0.6$) than at $s=1$ ($\vrmsm
\simeq 0.68$).

The higher velocities of the semipoles should make collisions more
frequent than those between monopoles and antimonopoles.  However, as
observed in the Introduction, semipole collisions need not result in
annihilation, and so the higher velocities do not necessarily result
in a lower monopole density.

We interpret the difference $\vrel^2 = \vrmsm^2 - \vrms^2$ as the mean
square relative velocity of the monopoles and semipoles along the
string.  One can estimate that, for semipoles, $\vrel \simeq 0.3$,
while there is little evidence for relative motion of monopoles.

\section{Conclusions}

We have carried out simulations of non-Abelian cosmic strings, formed
by the symmetry-breaking scheme SU(2)$\to Z_2$ by two adjoint scalar
fields.  This theory has classical solutions which can be interpreted
as 't Hooft-Polyakov monopoles or semipoles~\cite{Hindmarsh:2016lhy}
threaded by non-Abelian strings.  We observe the formation of cosmic
necklaces, consisting of networks of strings and monopoles or
semipoles.

Our simulations were carried out in a cosmological background
corresponding to a radiation dominated era, and also one with half the
expansion rate of a radiation-dominated universe, testing the effect
of the expansion rate.  We performed simulations both with the true
expanding universe equations of motion, and allowing the cores of the
topological defects to grow with the expansion of the universe.  Core
growth has been shown not to significantly affect the dynamics of
strings \cite{Bevis:2006mj,Bevis:2010gj,Daverio:2015nva}, but its
effect on the dynamics of necklaces is important to check.

In all cases, our numerical results are consistent with the evolution
towards a scaling network of necklaces, with both the density of
strings and the density of monopoles proportional to $t^{-2}$.  We
obtain scaling with or without core growth, giving confidence that
scaling is a robust feature of a necklace network. A necklace network
should therefore contribute a constant fraction to the energy density
of the universe.

We observe that the number of monopoles per unit comoving length of
string $n$ changes little from its value at the formation of the
string network: monopole annihilation on the string is therefore not
as efficient as envisaged in Ref.~\cite{BlancoPillado:2007zr}, and the
average comoving separation of monopoles along the string $d = 1/n$
remains approximately constant.  The monopole to string density ratio
$r$ therefore decreases in inverse proportion to the scale factor, and
does not increase as proposed in Ref.~\cite{Berezinsky:1997td}.  The
RMS monopole velocity is close to the RMS string velocity, implying
that the monopoles have no significant motion along the string.  In
particular, the suggestion that the monopole RMS velocity should be
50\% larger than the string RMS velocity \cite{BlancoPillado:2007zr},
due to the extra degree of freedom or motion, is not supported.

The number per unit comoving length of semipoles is also approximately
constant in the simulations with core growth, but grows slightly in
the simulations using the true equations of motion.  We do not have
large enough dynamic range to establish whether this is a finite
volume effect.  The semipole RMS velocity is higher than the string
RMS velocity, indicating some relative motion of the semipoles along
the string.  Annihilation is still inefficient despite the relative
motion, indicating that repulsion between semipoles is an important
factor in the dynamics.

In the special case where the strings carry a supercurrent, the
comoving distance between points where the $\Phi_1$ field vanishes $d$
also stays approximately constant.  The supercurrent along the string
can be estimated as $j \sim 1/ad$, where $a$ is the scale factor, and
should therefore decrease.  This suggests that current is lost from
shrinking loops of string, which would tend to prevent the formation
of cosmologically disastrous stable string loops
\cite{Ostriker:1986xc,Copeland:1987th,Davis:1988ij}.

We are restricted to simulating necklace configurations with $r \sim
1$, so we are not able to fully test the robustness of the of the
constant comoving $d$ scaling regime.  Nonetheless, we find it
interesting to explore the consequences as it was not anticipated in
previous dynamical modelling, which envisaged that $d$ would either
shrink to the string width ~\cite{Berezinsky:1997td}, or grow with the
horizon size \cite{BlancoPillado:2007zr}.  The absence of an
significant relative velocity between monopoles and strings indicates
that monopoles are dragged around by the strings, independent of the
ratio of the energy scales.  The average string separation is of order
the conformal time $t$, which means that loops of string shrink and
annihilate on that timescale. We infer that the main monopole
annihilation channel is though collisions on shrinking loops of
string.

As argued in \cite{Hindmarsh:2016lhy}, semipoles and monopoles are
generic on strings in GUT models.  It is interesting to consider their
observational implications.  As usual with strings, one must
extrapolate the results of numerical simulations to a much larger
ratio of the horizon size to the string width, and it is possible that
subtle effects change the scaling of the network.  It is clear in our
simulations that, just as with Abelian Higgs strings, our SU(2)
strings lose energy efficiently into Higgs and gauge radiation.
However, the process that causes the strings to emit radiation of
massive Higgs and gauge fields is not well understood, and it may not
be efficient over the huge range of scales between today's horizon
size and the width of a GUT string. In this case, a necklace would end
up behaving like ideal Nambu-Goto strings connecting massive
particles, as assumed in \cite{Berezinsky:1997td} and
\cite{BlancoPillado:2007zr}.

In the case where field radiation is efficient, there is little
difference between a network of GUT strings with monopoles or
semipoles and an Abelian Higgs string network.  The network length
scale grows in proportion to the horizon, and its energy density
remains a constant fraction of the total. The energy is lost to
massive particles, which (if coupled to the Standard Model) will show
up in the diffuse $\ga$-ray background.  Current observations from
Fermi-LAT indicate that the mass per unit length in Planck units
$G\mu$ is bounded above by $3 \times 10^{-11} f^{-1}_\text{SM}$, where
$f_\text{SM}$ is the fraction of the strings energy ending up in
$\ga$-rays~\cite{Mota:2014uka}. This fraction is likely to be close to
unity in a GUT theory, and so such strings are essentially ruled out,
as observed some time ago \cite{Vincent:1997cx}. However, strings in a
hidden sector are subject only to constraints from the Cosmic
Microwave
Background~\cite{Moss:2014cra,Charnock:2016nzm,Lizarraga:2016onn},
which are $G\mu \lesssim 10^{-7}$.

In the case where the string dynamics eventually changes over to
Nambu-Goto, the difference between a necklace network and an ordinary
cosmic string network is more dramatic with our new picture that the
comoving distance between monopoles remains approximately constant
from the time the strings formed.  For GUT scale strings forming along
with the monopoles, this is bounded above by the horizon distance at
the GUT temperature, or a few metres today.  Even if the scale of the
U(1) symmetry-breaking is as low as a TeV, this distance is
O($10^{12}$) m today, a factor $10^{-14}$ smaller than the horizon
size.  When horizon-size string loops are chopped off the long string
network, they will therefore have a large number of monopoles on them.
Numerical investigations indicate \cite{Siemens:2000ty} that such
string loops do not have periodic non-self-intersecting solutions.  We
can therefore expect them to quickly chop themselves up into smaller
and smaller loops, some of which will be free of monopoles and find
stable periodic non-self-intersecting trajectories.  In this case, the
typical loop size for a GUT scale string would be a few metres rather
than the horizon size.  Hence, the tight bounds on the Nambu-Goto
string tension from msec pulsar timing obtained by the European Pulsar
Timing Array \cite{Lentati:2015qwp} and NANOGrav
\cite{Arzoumanian:2015liz} would be avoided, as the gravitational
waves would be at frequencies inaccessible to direct observation.

\begin{acknowledgments}
We acknowledge fruitful discussions with Jarkko J\"arvel\"a during the
initial stages of this project. Our simulations made use of the COSMOS
Consortium supercomputer (within the DiRAC Facility jointly funded by
STFC and the Large Facilities Capital Fund of BIS). DJW was supported
by the People Programme (Marie Sk{\l}odowska-Curie actions) of the
European Union Seventh Framework Programme (FP7/2007-2013) under grant
agreement number PIEF-GA-2013-629425.  MH acknowledges support from
the Science and Technology Facilities Council (grant number
ST/L000504/1).
\end{acknowledgments}

\appendix
\section{Equations of motion on the lattice}
\label{app:lattice}

We write the adjoint Higgs field as $\Phi_n = \phi_n^a \ta^a$, with
$n=1,2$. The link variables for the gauge field are $U_\mu = u^0 \, 1
+ i\, u^a \sigma^a $, with $u^a \in \mathbb{R}$.

\begin{widetext}
The $d=4$ continuum action in a FLRW background with scale factor $a$
and $s=1$ is
\begin{equation}
S
 = \int d^4 x \; \left( -\frac{1}{4}  F_{\mu\nu}^a F^{\mu\nu a} + a^2 \sum_n \mathrm{Tr} [D_\mu,\Phi_n][D^\mu,\Phi_n] - a^4V(\Phi_1, \Phi_2) \right)
\label{e:SU2HLagFLRW}
\end{equation}
where indices are raised with the Minkowski metric $\eta_{\mu\nu} =
\diag(1,-1,-1,-1)_{\mu\nu}$.

As discussed both in Section~\ref{s:LatImp} and in
Appendix~\ref{app:expanding} below, in order to mitigate the shrinking
of the string and monopole cores in comoving coordinates one can allow
the coupling constants and mass parameters to become time-dependent,
\begin{equation}
    m_{1,2}^2 \to \frac{m_{1,2}^2}{a^{2(1-s)}}, \quad \lambda \to \frac{\lambda}{a^{2(1-s)}}, \quad g \to \frac{g}{a^{2(1-s)}}.
\end{equation}
The physical string and monopole core widths can be set to grow by
choosing $ s<1 $, with $s=0$ maintaining constant comoving core
widths.  This completely avoids the possibility of the topological
defects shrinking in size below the lattice spacing, although the
effect on their dynamics must be checked. In this paper we have used
$s=1$ and $s=0$ only.

With this in mind, we take the lattice action to be
\begin{align}
S[U,\Phi] & = \frac{4}{g^2 a^{2(s-1)}} \sum_{x; \; i} \left[1 - \frac{1}{2}\text{Tr} \;
  U_{0i}(x)\right] - \frac{4}{g^2 a^{2(s-1)}} \sum_{x; \; i < j} \left[1 - \frac{1}{2}\text{Tr} \;
  U_{ij}(x)\right]  \nonumber \\
& \quad + \sum_{x; \; n} a^2 \mathrm{Tr} [D_0,\Phi_n][D_0,\Phi_n] - \sum_{x; \; i,n} a^2 \mathrm{Tr} [D_i,\Phi_n][D_i,\Phi_n]  - \sum_{x} a^{4} V(\Phi_1, \Phi_2)
\end{align}
with unit comoving lattice spacing and scale factor $a$. The covariant
derivative is
\begin{equation}
[D_\mu,\Phi_n](x) = U_\mu(x) \Phi_n(x+\hat{\mu}) U_\mu^\dag(x) - \Phi_n(x).
\end{equation}
In the temporal gauge $U_0(x) = 1$,
\begin{align}
  S[U,\Phi] & = \frac{4}{g^2 a^{2(s-1)}} \sum_{x; \; i} \left[1 - \frac{1}{2}\text{Tr} \;
  U_{0i}(x)\right] - \frac{4}{g^2 a^{2(s-1)}} \sum_{x; \; i < j} \left[1 - \frac{1}{2}\text{Tr} \;
  U_{ij}(x)\right] \nonumber \\
& \quad + \sum_{x; \; n} a^2 \mathrm{Tr} \; \dot\Phi_n^2  - \sum_{x; \; i,n} a^2 \left[2 \, \mathrm{Tr} \, \Phi_n^2 - 2 \mathrm{Tr}\;
    \Phi_n(x)U_i(x)\Phi_n(x+\hat{\imath}) U_i^\dag (x) \right]  - \sum_{x} a^{4} V(\Phi_1, \Phi_2)
\end{align}
and after a Legendre transformation, the full Hamiltonian is
\begin{multline}
  H(t) = \frac{1}{2g^2 a^{2(s-1)}} \sum_{x,i,a} \ep_i^a(x,t)^2 + \frac{1}{2} a^2 \sum_{x; \; n,a} \; \pi_n^a(x,t)^2 
 + \frac{4}{g^2 a^{2(s-1)}} \sum_{x; \; i<j} \left( 1- \frac{1}{2} \text{Tr}
\; U_{i j} (x,t) \right) \\ 
- a^2 \sum_{x; \; i, n} 2 \; \mathrm{Tr}\;
\Phi_n(x)U_i(x)\Phi_n(x+\hat{\imath}) U_i^\dag (x)  
+ a^2 \sum_{x,n} 6 \, \mathrm{Tr} \; \Phi_n^2 + a^{4} \sum_{x} V(\Phi_1, \Phi_2)
\label{eq:firstham}.
\end{multline}
The equations of motion on the lattice are (recalling that we use the
label $a$ for elements of the Lie algebra and $n$ to label separate
fields)
\begin{align}
  g^2 a^{2(s-1)}\frac{\pa}{\pa t}\left( \frac{{\ep}_i^a (\mathbf{x},t)}{g^2 a^{2(s-1)}} \right)  &= -\sum_{j\neq i} \, 
\mathrm{Tr} \left\{ i \sigma^a U_{ij} (\mathbf{x},t) \right\}   
\notag \\
& \quad + g^2 a^{2s} \sum_{n} \Big[ - i\; \mathrm{Tr} \, \left\{\Phi_n(\mathbf{x},t) \sigma^a U_i(\mathbf{x},t)
\Phi_n(\mathbf{x} + \hat{\imath},t) U_i^\dag (\mathbf{x},t)  \right\} \notag \label{eq:eom:edot} \\
& \qquad +  i\; \mathrm{Tr} \, \left\{ \Phi_n(\mathbf{x},t) U_i(\mathbf{x},t) \Phi_n(\mathbf{x}+\hat{\imath},t) U_i^\dag
(\mathbf{x},t) \sigma^a \right\} \Big] \\
\dot{U}_i(\mathbf{x},t)
& =  - i \ep_i(\mathbf{x},t)
U_i(\mathbf{x},t)  \label{eq:eom:udot}  \\
\frac{1}{a^2}\frac{\pa}{\pa t}\left(a^2{\pi}_n^a(\mathbf{x},t)\right)  & = 6 \phi_n^a + a^{2} \frac{\partial V(\Phi_1,\Phi_2)}{\partial \phi_n^a}   -   \sum_j \mathrm{Tr} \left[ {\sigma^a}
  U_j(\mathbf{x},t) \Phi_n(\mathbf{x}+\hat{\jmath},t) U^\dag_j(\mathbf{x},t)\right] \notag \\
& \qquad -  \sum_j \mathrm{Tr} \left[ {\sigma^a}
  U_j^\dag(\mathbf{x}-\hat{\jmath},t) \Phi_n(\mathbf{x}-\hat{\jmath},t) U_j(\mathbf{x}-\hat{\jmath},t)\right]
\label{eq:eom:pidot} \\
\dot{\phi}^a_n (\mathbf{x},t) &= \pi^a_n (\mathbf{x},t)
\end{align}
where, for example
\begin{equation}
    \frac{\partial V(\Phi_1,\Phi_2)}{\partial \phi_1^a} = \frac{1}{a^{2(1-s)}} \left[ m_1^2  \phi_1^a+ 2 \lambda  (\mathrm{Tr}\Phi_1^2) \phi_1^a + \kappa  (\mathrm{Tr}\Phi_1\Phi_2) \phi_2^a \right],
\end{equation}
and similarly for $\phi_2^a$.

The Gauss law is
\begin{equation}
  \label{eq:gausslaw}
  G(x) = \sum_i \mathrm{Re} \; \mathrm{Tr} \; \sigma^a \,
  \left(\ep_i(x) 
  - U_i^\dag (x-\hat{\imath}) \ep_\nu (x-\hat{\imath})
  U_i(x-\hat{\imath}) \right) 
  - \rho(x) = 0
\end{equation}
where the scalar charge density $\rho(x)$ is
\begin{equation}
  \label{eq:charge}
  \rho(x) = 2g^2 a^{2s} \sum_n \; \mathrm{Tr} \; \sigma^a (\Pi_n \Phi_n -
  \Phi_n \Pi_n).
\end{equation}

\subsection{Remarks on the numerical implementation}

The implicit damping terms in Eqs.~(\ref{eq:eom:edot},
\ref{eq:eom:pidot}) are handled by a method of the Crank-Nicolson
type~\cite{Crank1996}. For Eq.~(\ref{eq:eom:edot}), let us write the
right hand side as $F\{U_i(x,t),\Phi(x,t)\}$. Then we have
\begin{equation}
  {\dot \ep}_i^a + 2 (1-s) \frac{\dot a}{a} {\ep}_i^a = F\{U_i(x,t), \Phi(x,t) \},
\end{equation}
which can be discretised as
\begin{equation}
  \frac{{\ep}_i^a ( t +\delta t/2) - {\ep}_i^a ( t - \delta t/2)}{\delta t} \\
  + (1-s) \frac{a(t + \delta t/2) - a(t-\delta t/2)}{\delta t \, a(t + \delta t/2)} \left[  {\ep}_i^a ( t +\delta t/2) + {\ep}_i^a ( t -\delta t/2) \right] = F\{ U_i(x,t) \Phi(x,t) \}.
\end{equation}
A similar expression can then be found for Eq.~(\ref{eq:eom:pidot}).

The gauge field evolution equation~(\ref{eq:eom:udot}) can be solved
to give
\begin{equation}
 U_i(x,t+\delta t) = \exp \left\{- i\frac{\sigma^j}{2} \ep_i^j(x,t+\delta t/2) \; \delta t
\right\}U_i(x,t) .
\end{equation}

We carry out a period of cooling prior to the core growth (in $s=1$)
or Hubble (in $s=0$) phases. The coupling constants and scale factor
are kept constant, and damping terms $\sigma \ep$ and $\sigma \pi_n^a$
are added to Eqs.~(\ref{eq:eom:edot}) and~(\ref{eq:eom:pidot})
respectively. This particular choice preserves the Gauss law. We
needed to use a very small timestep $\delta t=0.025$ during this short
cooling phase.

\end{widetext}

\section{Simulation in an expanding universe: core growth}
\label{app:expanding}

In the comoving coordinates of the lattice, the cores of defects
shrink as $a^{-1}$, where $a$ is the cosmological scale factor, as the
simulation proceeds. If the lattice resolution is to be sufficient to
resolve the core widths at the end of the simulation, the core would
be larger than the simulation box size $L$ in the initial conditions.
The core widths are also related to the time for the fields to relax
to their minima, making the production of a defect network from random
initial conditions hard to achieve.

To address the problem we scale the parameters of the theory by powers
of $a^{1-s}$, with $0 \le s \le 1$, as in Eq.~(\ref{e:ModHam}).  This
makes the comoving widths of the strings and monopoles $\Swid$ and
$\Mwid$ proportional to $a^{-s}$.  At $s=0$, the comoving width is
constant.  The properties of Abelian Higgs string networks are largely
independent of $s$, as they are controlled by the string tension,
which is invariant under this
scaling~\cite{Bevis:2006mj,Bevis:2010gj,Daverio:2015nva}.

\begin{figure}[htbp]
   \centering
   \includegraphics[width=0.5\textwidth]{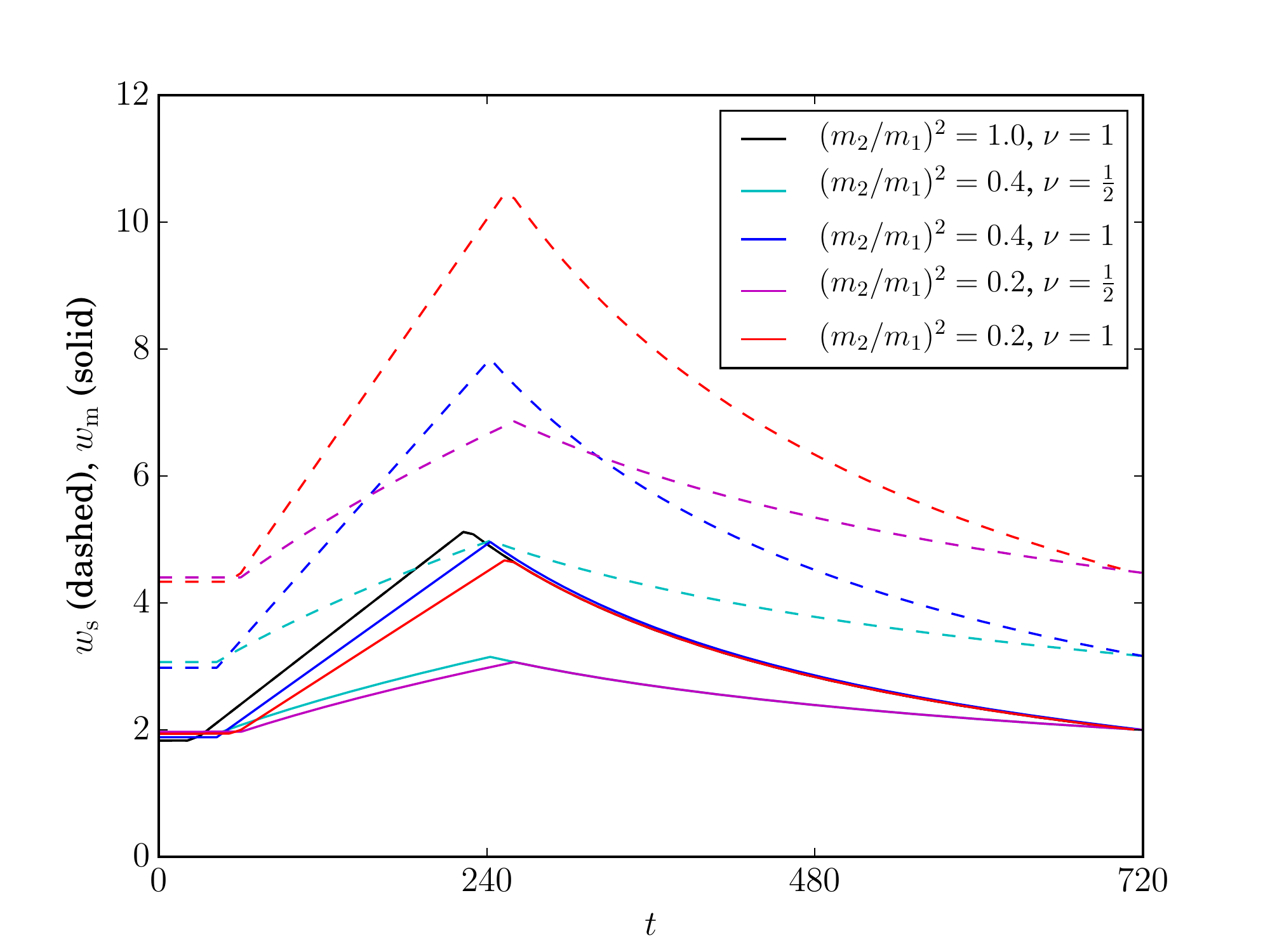} 
   \caption{Comoving core widths of monopoles ($\Mwid$, solid) and
     strings ($\Swid$, dashed) in the $s=1$ simulations. The core
     widths are defined as $\Mwid = (am_1)^{-1}$ and $\Swid =
     (am_2)^{-1}$, where $a$ is the scale factor. 
   \label{f:ComCorWid}}
\end{figure}

In order to simulate at $s=1$, we control $s$ through the simulation
so that the core width is small in the initial conditions, and grows
to meet the physical core width at a controllable time $\tCG$.  The
core widths of the strings and monopoles in our $s=1$ simulations are
plotted in Fig.~\ref{f:ComCorWid}.

\section{Projectors, magnetic charge, and winding number}
\label{app:projectors}

In this Appendix, we follow Ref.~\cite{Davis:2000kv} in denoting the
two Higgs fields in the adjoint representation by $\Phi$ and
$\chi$. We will assume that $\Phi=\Phi_1$ forms the 't Hooft-Polyakov
monopoles, while $\chi=\Phi_2$ is responsible for the strings.

We define the projectors $\Pi_\pm = \frac{1}{2} (1 \pm \hat{\Phi})$,
where $\hat{\Phi} = \Phi\sqrt{2/\mathrm{Tr}\, \Phi^2}$ (similarly,
$\hat{\chi} = \chi\sqrt{2/\mathrm{Tr}\, \chi^2}$).

\subsection{Magnetic charge}
For the time being, in this section we will return to the full
4-dimensional theory.

In the symmetry broken phase, a residual U(1) symmetry persists. We
can derive link variables $u_\mu$ corresponding to this smaller gauge
group~\cite{Davis:2000kv,Edwards:2009bw},
\begin{equation}
u_\mu(x) = \Pi_+(x) U_\mu(x) \Pi_+(x+\hat{\mu});
\end{equation}
these can be shown to transform like the Abelian gauge field. The
corresponding Abelian field strength tensor is
\begin{equation}
A_{\mu\nu} = \mathrm{arg} \; \mathrm{Tr} \; u_\mu(x) u_\nu(x+\hat{\mu}) u_\mu^\dag(x+\hat{\nu}) u_\nu^\dag(x)
\end{equation}
and, with the correct factors of the coupling to give a continuum
electromagnetic field,
\begin{align}
\alpha_{\mu\nu} & = \frac{2}{g} A_{\mu\nu} \\
& = \frac{2}{g} \; \mathrm{arg} \; \mathrm{Tr}\; u_\mu(x) u_\nu(x+\hat{\mu}) u_\mu^\dag(x+\hat{\nu}) u_\nu^\dag(x)
\end{align}
and, finally, the expression for the lattice magnetic field
\begin{equation}
B_i = \frac{1}{2} \epsilon_{ijk} \alpha_{jk}.
\end{equation}

The symmetry breaking phase transitions studied in this work allow the
creation of magnetic charge. On the lattice, the projected Gauss law
for the magnetic field takes the form
\begin{equation}
  \sum_{i=1}^3 [B_i(x+\hat{\imath}) - B_i(x)] = \rho_\mathrm{M}(x)
  = \frac{4\pi N}{g}
\end{equation}
where $N$ is an integer. It is important to note that the magnetic
charge is quantised and localised within lattice cells.

\subsection{Winding number}

The above section yielded $\alpha_{\mu\nu}$, the analogue of the
Abelian gauge field, and hence $E_i$ and $B_i$. To measure the winding
number directly~\cite{Davis:2000kv}, we also need to find the
equivalent of the Abelian Higgs field, as its phase angle appears in
the definition of winding number.

The difference in phase angle for the residual Higgs field at
neighbouring lattice sites $(x, x+\hat\imath)$ is then
\begin{multline}
\delta_i(x) = \mathrm{arg} \; \mathrm{Tr} \; \big[ \hat{\chi}(x) \Pi_-(x) U_i(x) \Pi_-(x+\hat{\imath}) \\ 
\hat{\chi}(x +\hat{\imath}) \Pi_+(x+\hat{\imath}) U_i^\dag(x) \Pi_+(x) \big].
\end{multline}
The winding number through a plaquette is then
\begin{multline}
\label{eq:plaqwinding}
Y_{ij}(x) = \delta_i(x) + \delta_j(x+\hat{\imath}) \\ 
- \delta_i(x+\hat{\jmath}) - \delta_j(x) - 2 A_{ij}(x)
\end{multline}
which is gauge-invariant.

We then approximate the string length $L$ in the system by the total
string winding through all plaquettes,
\begin{equation}
\sum_{x; i<j} Y_{ij}(x) = 2\pi L.
\end{equation}

\bibliography{strings}

\begin{thebibliography}{53}%
\makeatletter
\providecommand \@ifxundefined [1]{%
 \@ifx{#1\undefined}
}%
\providecommand \@ifnum [1]{%
 \ifnum #1\expandafter \@firstoftwo
 \else \expandafter \@secondoftwo
 \fi
}%
\providecommand \@ifx [1]{%
 \ifx #1\expandafter \@firstoftwo
 \else \expandafter \@secondoftwo
 \fi
}%
\providecommand \natexlab [1]{#1}%
\providecommand \enquote  [1]{``#1''}%
\providecommand \bibnamefont  [1]{#1}%
\providecommand \bibfnamefont [1]{#1}%
\providecommand \citenamefont [1]{#1}%
\providecommand \href@noop [0]{\@secondoftwo}%
\providecommand \href [0]{\begingroup \@sanitize@url \@href}%
\providecommand \@href[1]{\@@startlink{#1}\@@href}%
\providecommand \@@href[1]{\endgroup#1\@@endlink}%
\providecommand \@sanitize@url [0]{\catcode `\\12\catcode `\$12\catcode
  `\&12\catcode `\#12\catcode `\^12\catcode `\_12\catcode `\%12\relax}%
\providecommand \@@startlink[1]{}%
\providecommand \@@endlink[0]{}%
\providecommand \url  [0]{\begingroup\@sanitize@url \@url }%
\providecommand \@url [1]{\endgroup\@href {#1}{\urlprefix }}%
\providecommand \urlprefix  [0]{URL }%
\providecommand \Eprint [0]{\href }%
\providecommand \doibase [0]{http://dx.doi.org/}%
\providecommand \selectlanguage [0]{\@gobble}%
\providecommand \bibinfo  [0]{\@secondoftwo}%
\providecommand \bibfield  [0]{\@secondoftwo}%
\providecommand \translation [1]{[#1]}%
\providecommand \BibitemOpen [0]{}%
\providecommand \bibitemStop [0]{}%
\providecommand \bibitemNoStop [0]{.\EOS\space}%
\providecommand \EOS [0]{\spacefactor3000\relax}%
\providecommand \BibitemShut  [1]{\csname bibitem#1\endcsname}%
\let\auto@bib@innerbib\@empty
%</preamble>
\bibitem [{\citenamefont {Kibble}(1976)}]{Kibble:1976sj}%
  \BibitemOpen
  \bibfield  {author} {\bibinfo {author} {\bibfnamefont {T.}~\bibnamefont
  {Kibble}},\ }\href {\doibase 10.1088/0305-4470/9/8/029} {\bibfield  {journal}
  {\bibinfo  {journal} {J.Phys.}\ }\textbf {\bibinfo {volume} {A9}},\ \bibinfo
  {pages} {1387} (\bibinfo {year} {1976})}\BibitemShut {NoStop}%
%%CITATION = ICTP/75/5 ETC.;%%
\bibitem [{\citenamefont {Hindmarsh}\ and\ \citenamefont
  {Kibble}(1995)}]{Hindmarsh:1994re}%
  \BibitemOpen
  \bibfield  {author} {\bibinfo {author} {\bibfnamefont {M.}~\bibnamefont
  {Hindmarsh}}\ and\ \bibinfo {author} {\bibfnamefont {T.}~\bibnamefont
  {Kibble}},\ }\href {\doibase 10.1088/0034-4885/58/5/001} {\bibfield
  {journal} {\bibinfo  {journal} {Rept.Prog.Phys.}\ }\textbf {\bibinfo {volume}
  {58}},\ \bibinfo {pages} {477} (\bibinfo {year} {1995})},\ \Eprint
  {http://arxiv.org/abs/hep-ph/9411342} {arXiv:hep-ph/9411342 [hep-ph]}
  \BibitemShut {NoStop}%
%%CITATION = HEP-PH/9411342;%%
\bibitem [{\citenamefont {Vilenkin}\ and\ \citenamefont
  {Shellard}(2000)}]{Vilenkin:2000jqa}%
  \BibitemOpen
  \bibfield  {author} {\bibinfo {author} {\bibfnamefont {A.}~\bibnamefont
  {Vilenkin}}\ and\ \bibinfo {author} {\bibfnamefont {E.~P.~S.}\ \bibnamefont
  {Shellard}},\ }\href
  {http://www.cambridge.org/mw/academic/subjects/physics/theoretical-physics-and-mathematical-physics/cosmic-strings-and-other-topological-defects?format=PB}
  {\emph {\bibinfo {title} {{Cosmic Strings and Other Topological Defects}}}}\
  (\bibinfo  {publisher} {Cambridge University Press},\ \bibinfo {year}
  {2000})\BibitemShut {NoStop}%
%%CITATION = INSPIRE-1384873;%%
\bibitem [{\citenamefont {Copeland}\ \emph {et~al.}(2011)\citenamefont
  {Copeland}, \citenamefont {Pogosian},\ and\ \citenamefont
  {Vachaspati}}]{Copeland:2011dx}%
  \BibitemOpen
  \bibfield  {author} {\bibinfo {author} {\bibfnamefont {E.~J.}\ \bibnamefont
  {Copeland}}, \bibinfo {author} {\bibfnamefont {L.}~\bibnamefont {Pogosian}},
  \ and\ \bibinfo {author} {\bibfnamefont {T.}~\bibnamefont {Vachaspati}},\
  }\href {\doibase 10.1088/0264-9381/28/20/204009} {\bibfield  {journal}
  {\bibinfo  {journal} {Class.Quant.Grav.}\ }\textbf {\bibinfo {volume} {28}},\
  \bibinfo {pages} {204009} (\bibinfo {year} {2011})},\ \Eprint
  {http://arxiv.org/abs/1105.0207} {arXiv:1105.0207 [hep-th]} \BibitemShut
  {NoStop}%
%%CITATION = ARXIV:1105.0207;%%
\bibitem [{\citenamefont {Hindmarsh}(2011)}]{Hindmarsh:2011qj}%
  \BibitemOpen
  \bibfield  {author} {\bibinfo {author} {\bibfnamefont {M.}~\bibnamefont
  {Hindmarsh}},\ }\href {\doibase 10.1143/PTPS.190.197} {\bibfield  {journal}
  {\bibinfo  {journal} {Prog.Theor.Phys.Suppl.}\ }\textbf {\bibinfo {volume}
  {190}},\ \bibinfo {pages} {197} (\bibinfo {year} {2011})},\ \Eprint
  {http://arxiv.org/abs/1106.0391} {arXiv:1106.0391 [astro-ph.CO]} \BibitemShut
  {NoStop}%
%%CITATION = ARXIV:1106.0391;%%
\bibitem [{\citenamefont {Nielsen}\ and\ \citenamefont
  {Olesen}(1973)}]{Nielsen:1973cs}%
  \BibitemOpen
  \bibfield  {author} {\bibinfo {author} {\bibfnamefont {H.~B.}\ \bibnamefont
  {Nielsen}}\ and\ \bibinfo {author} {\bibfnamefont {P.}~\bibnamefont
  {Olesen}},\ }\href {\doibase 10.1016/0550-3213(73)90350-7} {\bibfield
  {journal} {\bibinfo  {journal} {Nucl.Phys.}\ }\textbf {\bibinfo {volume}
  {B61}},\ \bibinfo {pages} {45} (\bibinfo {year} {1973})}\BibitemShut
  {NoStop}%
%%CITATION = NUPHA,B61,45;%%
\bibitem [{\citenamefont {Witten}(1985)}]{Witten:1985fp}%
  \BibitemOpen
  \bibfield  {author} {\bibinfo {author} {\bibfnamefont {E.}~\bibnamefont
  {Witten}},\ }\href {\doibase 10.1016/0370-2693(85)90540-4} {\bibfield
  {journal} {\bibinfo  {journal} {Phys.Lett.}\ }\textbf {\bibinfo {volume}
  {B153}},\ \bibinfo {pages} {243} (\bibinfo {year} {1985})}\BibitemShut
  {NoStop}%
%%CITATION = PHLTA,B153,243;%%
\bibitem [{\citenamefont {Sarangi}\ and\ \citenamefont
  {Tye}(2002)}]{Sarangi:2002yt}%
  \BibitemOpen
  \bibfield  {author} {\bibinfo {author} {\bibfnamefont {S.}~\bibnamefont
  {Sarangi}}\ and\ \bibinfo {author} {\bibfnamefont {S.~H.}\ \bibnamefont
  {Tye}},\ }\href {\doibase 10.1016/S0370-2693(02)01824-5} {\bibfield
  {journal} {\bibinfo  {journal} {Phys.Lett.}\ }\textbf {\bibinfo {volume}
  {B536}},\ \bibinfo {pages} {185} (\bibinfo {year} {2002})},\ \Eprint
  {http://arxiv.org/abs/hep-th/0204074} {arXiv:hep-th/0204074 [hep-th]}
  \BibitemShut {NoStop}%
%%CITATION = HEP-TH/0204074;%%
\bibitem [{\citenamefont {Copeland}\ \emph {et~al.}(2004)\citenamefont
  {Copeland}, \citenamefont {Myers},\ and\ \citenamefont
  {Polchinski}}]{Copeland:2003bj}%
  \BibitemOpen
  \bibfield  {author} {\bibinfo {author} {\bibfnamefont {E.~J.}\ \bibnamefont
  {Copeland}}, \bibinfo {author} {\bibfnamefont {R.~C.}\ \bibnamefont {Myers}},
  \ and\ \bibinfo {author} {\bibfnamefont {J.}~\bibnamefont {Polchinski}},\
  }\href {\doibase 10.1088/1126-6708/2004/06/013} {\bibfield  {journal}
  {\bibinfo  {journal} {JHEP}\ }\textbf {\bibinfo {volume} {0406}},\ \bibinfo
  {pages} {013} (\bibinfo {year} {2004})},\ \Eprint
  {http://arxiv.org/abs/hep-th/0312067} {arXiv:hep-th/0312067 [hep-th]}
  \BibitemShut {NoStop}%
%%CITATION = HEP-TH/0312067;%%
\bibitem [{\citenamefont {Lizarraga}\ and\ \citenamefont
  {Urrestilla}(2016)}]{Lizarraga:2016hpd}%
  \BibitemOpen
  \bibfield  {author} {\bibinfo {author} {\bibfnamefont {J.}~\bibnamefont
  {Lizarraga}}\ and\ \bibinfo {author} {\bibfnamefont {J.}~\bibnamefont
  {Urrestilla}},\ }\href {\doibase 10.1088/1475-7516/2016/04/053} {\bibfield
  {journal} {\bibinfo  {journal} {JCAP}\ }\textbf {\bibinfo {volume} {1604}},\
  \bibinfo {pages} {053} (\bibinfo {year} {2016})},\ \Eprint
  {http://arxiv.org/abs/1602.08014} {arXiv:1602.08014 [astro-ph.CO]}
  \BibitemShut {NoStop}%
%%CITATION = ARXIV:1602.08014;%%
\bibitem [{\citenamefont {Laguna}\ and\ \citenamefont
  {Matzner}(1989)}]{Laguna:1989hn}%
  \BibitemOpen
  \bibfield  {author} {\bibinfo {author} {\bibfnamefont {P.}~\bibnamefont
  {Laguna}}\ and\ \bibinfo {author} {\bibfnamefont {R.}~\bibnamefont
  {Matzner}},\ }\href {\doibase 10.1103/PhysRevLett.62.1948} {\bibfield
  {journal} {\bibinfo  {journal} {Phys.Rev.Lett.}\ }\textbf {\bibinfo {volume}
  {62}},\ \bibinfo {pages} {1948} (\bibinfo {year} {1989})}\BibitemShut
  {NoStop}%
%%CITATION = PRLTA,62,1948;%%
\bibitem [{\citenamefont {Vincent}\ \emph {et~al.}(1998)\citenamefont
  {Vincent}, \citenamefont {Antunes},\ and\ \citenamefont
  {Hindmarsh}}]{Vincent:1997cx}%
  \BibitemOpen
  \bibfield  {author} {\bibinfo {author} {\bibfnamefont {G.}~\bibnamefont
  {Vincent}}, \bibinfo {author} {\bibfnamefont {N.~D.}\ \bibnamefont
  {Antunes}}, \ and\ \bibinfo {author} {\bibfnamefont {M.}~\bibnamefont
  {Hindmarsh}},\ }\href {\doibase 10.1103/PhysRevLett.80.2277} {\bibfield
  {journal} {\bibinfo  {journal} {Phys.Rev.Lett.}\ }\textbf {\bibinfo {volume}
  {80}},\ \bibinfo {pages} {2277} (\bibinfo {year} {1998})},\ \Eprint
  {http://arxiv.org/abs/hep-ph/9708427} {arXiv:hep-ph/9708427 [hep-ph]}
  \BibitemShut {NoStop}%
%%CITATION = HEP-PH/9708427;%%
\bibitem [{\citenamefont {Moore}\ \emph {et~al.}(2002)\citenamefont {Moore},
  \citenamefont {Shellard},\ and\ \citenamefont {Martins}}]{Moore:2001px}%
  \BibitemOpen
  \bibfield  {author} {\bibinfo {author} {\bibfnamefont {J.}~\bibnamefont
  {Moore}}, \bibinfo {author} {\bibfnamefont {E.}~\bibnamefont {Shellard}}, \
  and\ \bibinfo {author} {\bibfnamefont {C.}~\bibnamefont {Martins}},\ }\href
  {\doibase 10.1103/PhysRevD.65.023503} {\bibfield  {journal} {\bibinfo
  {journal} {Phys.Rev.}\ }\textbf {\bibinfo {volume} {D65}},\ \bibinfo {pages}
  {023503} (\bibinfo {year} {2002})},\ \Eprint
  {http://arxiv.org/abs/hep-ph/0107171} {arXiv:hep-ph/0107171 [hep-ph]}
  \BibitemShut {NoStop}%
%%CITATION = HEP-PH/0107171;%%
\bibitem [{\citenamefont {Bevis}\ \emph {et~al.}(2007)\citenamefont {Bevis},
  \citenamefont {Hindmarsh}, \citenamefont {Kunz},\ and\ \citenamefont
  {Urrestilla}}]{Bevis:2006mj}%
  \BibitemOpen
  \bibfield  {author} {\bibinfo {author} {\bibfnamefont {N.}~\bibnamefont
  {Bevis}}, \bibinfo {author} {\bibfnamefont {M.}~\bibnamefont {Hindmarsh}},
  \bibinfo {author} {\bibfnamefont {M.}~\bibnamefont {Kunz}}, \ and\ \bibinfo
  {author} {\bibfnamefont {J.}~\bibnamefont {Urrestilla}},\ }\href {\doibase
  10.1103/PhysRevD.75.065015} {\bibfield  {journal} {\bibinfo  {journal}
  {Phys.Rev.}\ }\textbf {\bibinfo {volume} {D75}},\ \bibinfo {pages} {065015}
  (\bibinfo {year} {2007})},\ \Eprint {http://arxiv.org/abs/astro-ph/0605018}
  {arXiv:astro-ph/0605018 [astro-ph]} \BibitemShut {NoStop}%
%%CITATION = ASTRO-PH/0605018;%%
\bibitem [{\citenamefont {Bevis}\ \emph {et~al.}(2010)\citenamefont {Bevis},
  \citenamefont {Hindmarsh}, \citenamefont {Kunz},\ and\ \citenamefont
  {Urrestilla}}]{Bevis:2010gj}%
  \BibitemOpen
  \bibfield  {author} {\bibinfo {author} {\bibfnamefont {N.}~\bibnamefont
  {Bevis}}, \bibinfo {author} {\bibfnamefont {M.}~\bibnamefont {Hindmarsh}},
  \bibinfo {author} {\bibfnamefont {M.}~\bibnamefont {Kunz}}, \ and\ \bibinfo
  {author} {\bibfnamefont {J.}~\bibnamefont {Urrestilla}},\ }\href {\doibase
  10.1103/PhysRevD.82.065004} {\bibfield  {journal} {\bibinfo  {journal}
  {Phys.Rev.}\ }\textbf {\bibinfo {volume} {D82}},\ \bibinfo {pages} {065004}
  (\bibinfo {year} {2010})},\ \Eprint {http://arxiv.org/abs/1005.2663}
  {arXiv:1005.2663 [astro-ph.CO]} \BibitemShut {NoStop}%
%%CITATION = ARXIV:1005.2663;%%
\bibitem [{\citenamefont {Vachaspati}\ and\ \citenamefont
  {Vilenkin}(1987)}]{Vachaspati:1986cc}%
  \BibitemOpen
  \bibfield  {author} {\bibinfo {author} {\bibfnamefont {T.}~\bibnamefont
  {Vachaspati}}\ and\ \bibinfo {author} {\bibfnamefont {A.}~\bibnamefont
  {Vilenkin}},\ }\href {\doibase 10.1103/PhysRevD.35.1131} {\bibfield
  {journal} {\bibinfo  {journal} {Phys. Rev.}\ }\textbf {\bibinfo {volume}
  {D35}},\ \bibinfo {pages} {1131} (\bibinfo {year} {1987})}\BibitemShut
  {NoStop}%
%%CITATION = PHRVA,D35,1131;%%
\bibitem [{\citenamefont {Copeland}\ and\ \citenamefont
  {Saffin}(2005)}]{Copeland:2005cy}%
  \BibitemOpen
  \bibfield  {author} {\bibinfo {author} {\bibfnamefont {E.~J.}\ \bibnamefont
  {Copeland}}\ and\ \bibinfo {author} {\bibfnamefont {P.~M.}\ \bibnamefont
  {Saffin}},\ }\href {\doibase 10.1088/1126-6708/2005/11/023} {\bibfield
  {journal} {\bibinfo  {journal} {JHEP}\ }\textbf {\bibinfo {volume} {11}},\
  \bibinfo {pages} {023} (\bibinfo {year} {2005})},\ \Eprint
  {http://arxiv.org/abs/hep-th/0505110} {arXiv:hep-th/0505110 [hep-th]}
  \BibitemShut {NoStop}%
%%CITATION = HEP-TH/0505110;%%
\bibitem [{\citenamefont {Hindmarsh}\ and\ \citenamefont
  {Saffin}(2006)}]{Hindmarsh:2006qn}%
  \BibitemOpen
  \bibfield  {author} {\bibinfo {author} {\bibfnamefont {M.}~\bibnamefont
  {Hindmarsh}}\ and\ \bibinfo {author} {\bibfnamefont {P.}~\bibnamefont
  {Saffin}},\ }\href {\doibase 10.1088/1126-6708/2006/08/066} {\bibfield
  {journal} {\bibinfo  {journal} {JHEP}\ }\textbf {\bibinfo {volume} {0608}},\
  \bibinfo {pages} {066} (\bibinfo {year} {2006})},\ \Eprint
  {http://arxiv.org/abs/hep-th/0605014} {arXiv:hep-th/0605014 [hep-th]}
  \BibitemShut {NoStop}%
%%CITATION = HEP-TH/0605014;%%
\bibitem [{\citenamefont {Urrestilla}\ and\ \citenamefont
  {Vilenkin}(2008)}]{Urrestilla:2007yw}%
  \BibitemOpen
  \bibfield  {author} {\bibinfo {author} {\bibfnamefont {J.}~\bibnamefont
  {Urrestilla}}\ and\ \bibinfo {author} {\bibfnamefont {A.}~\bibnamefont
  {Vilenkin}},\ }\href {\doibase 10.1088/1126-6708/2008/02/037} {\bibfield
  {journal} {\bibinfo  {journal} {JHEP}\ }\textbf {\bibinfo {volume} {0802}},\
  \bibinfo {pages} {037} (\bibinfo {year} {2008})},\ \Eprint
  {http://arxiv.org/abs/0712.1146} {arXiv:0712.1146 [hep-th]} \BibitemShut
  {NoStop}%
%%CITATION = ARXIV:0712.1146;%%
\bibitem [{\citenamefont {Leblond}\ and\ \citenamefont
  {Wyman}(2007)}]{Leblond:2007tf}%
  \BibitemOpen
  \bibfield  {author} {\bibinfo {author} {\bibfnamefont {L.}~\bibnamefont
  {Leblond}}\ and\ \bibinfo {author} {\bibfnamefont {M.}~\bibnamefont
  {Wyman}},\ }\href {\doibase 10.1103/PhysRevD.75.123522} {\bibfield  {journal}
  {\bibinfo  {journal} {Phys. Rev.}\ }\textbf {\bibinfo {volume} {D75}},\
  \bibinfo {pages} {123522} (\bibinfo {year} {2007})},\ \Eprint
  {http://arxiv.org/abs/astro-ph/0701427} {arXiv:astro-ph/0701427 [astro-ph]}
  \BibitemShut {NoStop}%
%%CITATION = ASTRO-PH/0701427;%%
\bibitem [{\citenamefont {Martins}(2010)}]{Martins:2010ma}%
  \BibitemOpen
  \bibfield  {author} {\bibinfo {author} {\bibfnamefont {C.~J. A.~P.}\
  \bibnamefont {Martins}},\ }\href {\doibase 10.1103/PhysRevD.82.067301}
  {\bibfield  {journal} {\bibinfo  {journal} {Phys. Rev.}\ }\textbf {\bibinfo
  {volume} {D82}},\ \bibinfo {pages} {067301} (\bibinfo {year} {2010})},\
  \Eprint {http://arxiv.org/abs/1009.1707} {arXiv:1009.1707 [hep-ph]}
  \BibitemShut {NoStop}%
%%CITATION = ARXIV:1009.1707;%%
\bibitem [{\citenamefont {Hindmarsh}\ and\ \citenamefont
  {Kibble}(1985)}]{Hindmarsh:1985xc}%
  \BibitemOpen
  \bibfield  {author} {\bibinfo {author} {\bibfnamefont {M.}~\bibnamefont
  {Hindmarsh}}\ and\ \bibinfo {author} {\bibfnamefont {T.}~\bibnamefont
  {Kibble}},\ }\href {\doibase 10.1103/PhysRevLett.55.2398} {\bibfield
  {journal} {\bibinfo  {journal} {Phys.Rev.Lett.}\ }\textbf {\bibinfo {volume}
  {55}},\ \bibinfo {pages} {2398} (\bibinfo {year} {1985})}\BibitemShut
  {NoStop}%
%%CITATION = PRLTA,55,2398;%%
\bibitem [{\citenamefont {de~Vega}\ and\ \citenamefont
  {Schaposnik}(1986{\natexlab{a}})}]{deVega:1986eu}%
  \BibitemOpen
  \bibfield  {author} {\bibinfo {author} {\bibfnamefont {H.~J.}\ \bibnamefont
  {de~Vega}}\ and\ \bibinfo {author} {\bibfnamefont {F.~A.}\ \bibnamefont
  {Schaposnik}},\ }\href {\doibase 10.1103/PhysRevLett.56.2564} {\bibfield
  {journal} {\bibinfo  {journal} {Phys. Rev. Lett.}\ }\textbf {\bibinfo
  {volume} {56}},\ \bibinfo {pages} {2564} (\bibinfo {year}
  {1986}{\natexlab{a}})}\BibitemShut {NoStop}%
%%CITATION = PRLTA,56,2564;%%
\bibitem [{\citenamefont {de~Vega}\ and\ \citenamefont
  {Schaposnik}(1986{\natexlab{b}})}]{deVega:1986hm}%
  \BibitemOpen
  \bibfield  {author} {\bibinfo {author} {\bibfnamefont {H.~J.}\ \bibnamefont
  {de~Vega}}\ and\ \bibinfo {author} {\bibfnamefont {F.~A.}\ \bibnamefont
  {Schaposnik}},\ }\href {\doibase 10.1103/PhysRevD.34.3206} {\bibfield
  {journal} {\bibinfo  {journal} {Phys. Rev.}\ }\textbf {\bibinfo {volume}
  {D34}},\ \bibinfo {pages} {3206} (\bibinfo {year}
  {1986}{\natexlab{b}})}\BibitemShut {NoStop}%
%%CITATION = PHRVA,D34,3206;%%
\bibitem [{\citenamefont {Aryal}\ and\ \citenamefont
  {Everett}(1987)}]{Aryal:1987sn}%
  \BibitemOpen
  \bibfield  {author} {\bibinfo {author} {\bibfnamefont {M.}~\bibnamefont
  {Aryal}}\ and\ \bibinfo {author} {\bibfnamefont {A.~E.}\ \bibnamefont
  {Everett}},\ }\href {\doibase 10.1103/PhysRevD.35.3105} {\bibfield  {journal}
  {\bibinfo  {journal} {Phys. Rev.}\ }\textbf {\bibinfo {volume} {D35}},\
  \bibinfo {pages} {3105} (\bibinfo {year} {1987})}\BibitemShut {NoStop}%
%%CITATION = PHRVA,D35,3105;%%
\bibitem [{\citenamefont {Kibble}\ \emph {et~al.}(1982)\citenamefont {Kibble},
  \citenamefont {Lazarides},\ and\ \citenamefont {Shafi}}]{Kibble:1982ae}%
  \BibitemOpen
  \bibfield  {author} {\bibinfo {author} {\bibfnamefont {T.~W.~B.}\
  \bibnamefont {Kibble}}, \bibinfo {author} {\bibfnamefont {G.}~\bibnamefont
  {Lazarides}}, \ and\ \bibinfo {author} {\bibfnamefont {Q.}~\bibnamefont
  {Shafi}},\ }\href {\doibase 10.1016/0370-2693(82)90829-2} {\bibfield
  {journal} {\bibinfo  {journal} {Phys. Lett.}\ }\textbf {\bibinfo {volume}
  {B113}},\ \bibinfo {pages} {237} (\bibinfo {year} {1982})}\BibitemShut
  {NoStop}%
%%CITATION = PHLTA,B113,237;%%
\bibitem [{\citenamefont {Jeannerot}\ \emph {et~al.}(2003)\citenamefont
  {Jeannerot}, \citenamefont {Rocher},\ and\ \citenamefont
  {Sakellariadou}}]{Jeannerot:2003qv}%
  \BibitemOpen
  \bibfield  {author} {\bibinfo {author} {\bibfnamefont {R.}~\bibnamefont
  {Jeannerot}}, \bibinfo {author} {\bibfnamefont {J.}~\bibnamefont {Rocher}}, \
  and\ \bibinfo {author} {\bibfnamefont {M.}~\bibnamefont {Sakellariadou}},\
  }\href {\doibase 10.1103/PhysRevD.68.103514} {\bibfield  {journal} {\bibinfo
  {journal} {Phys. Rev.}\ }\textbf {\bibinfo {volume} {D68}},\ \bibinfo {pages}
  {103514} (\bibinfo {year} {2003})},\ \Eprint
  {http://arxiv.org/abs/hep-ph/0308134} {arXiv:hep-ph/0308134 [hep-ph]}
  \BibitemShut {NoStop}%
%%CITATION = HEP-PH/0308134;%%
\bibitem [{\citenamefont {'t~Hooft}(1974)}]{Hooft:1974qc}%
  \BibitemOpen
  \bibfield  {author} {\bibinfo {author} {\bibfnamefont {G.}~\bibnamefont
  {'t~Hooft}},\ }\href {\doibase 10.1016/0550-3213(74)90486-6} {\bibfield
  {journal} {\bibinfo  {journal} {Nucl. Phys.}\ }\textbf {\bibinfo {volume}
  {B79}},\ \bibinfo {pages} {276} (\bibinfo {year} {1974})}\BibitemShut
  {NoStop}%
%%CITATION = NUPHA,B79,276;%%
\bibitem [{\citenamefont {Polyakov}(1974)}]{Polyakov:1974ek}%
  \BibitemOpen
  \bibfield  {author} {\bibinfo {author} {\bibfnamefont {A.~M.}\ \bibnamefont
  {Polyakov}},\ }\href@noop {} {\bibfield  {journal} {\bibinfo  {journal} {JETP
  Lett.}\ }\textbf {\bibinfo {volume} {20}},\ \bibinfo {pages} {194} (\bibinfo
  {year} {1974})},\ \bibinfo {note} {[Pisma Zh. Eksp. Teor.
  Fiz.20,430(1974)]}\BibitemShut {NoStop}%
%%CITATION = JTPLA,20,194;%%
\bibitem [{\citenamefont {Berezinsky}\ and\ \citenamefont
  {Vilenkin}(1997)}]{Berezinsky:1997td}%
  \BibitemOpen
  \bibfield  {author} {\bibinfo {author} {\bibfnamefont {V.}~\bibnamefont
  {Berezinsky}}\ and\ \bibinfo {author} {\bibfnamefont {A.}~\bibnamefont
  {Vilenkin}},\ }\href {\doibase 10.1103/PhysRevLett.79.5202} {\bibfield
  {journal} {\bibinfo  {journal} {Phys.Rev.Lett.}\ }\textbf {\bibinfo {volume}
  {79}},\ \bibinfo {pages} {5202} (\bibinfo {year} {1997})},\ \Eprint
  {http://arxiv.org/abs/astro-ph/9704257} {arXiv:astro-ph/9704257 [astro-ph]}
  \BibitemShut {NoStop}%
%%CITATION = ASTRO-PH/9704257;%%
\bibitem [{\citenamefont {Kibble}\ and\ \citenamefont
  {Vachaspati}(2015)}]{Kibble:2015twa}%
  \BibitemOpen
  \bibfield  {author} {\bibinfo {author} {\bibfnamefont {T.~W.~B.}\
  \bibnamefont {Kibble}}\ and\ \bibinfo {author} {\bibfnamefont
  {T.}~\bibnamefont {Vachaspati}},\ }\href {\doibase
  10.1088/0954-3899/42/9/094002} {\bibfield  {journal} {\bibinfo  {journal} {J.
  Phys.}\ }\textbf {\bibinfo {volume} {G42}},\ \bibinfo {pages} {094002}
  (\bibinfo {year} {2015})},\ \Eprint {http://arxiv.org/abs/1506.02022}
  {arXiv:1506.02022 [astro-ph.CO]} \BibitemShut {NoStop}%
%%CITATION = ARXIV:1506.02022;%%
\bibitem [{\citenamefont {Hindmarsh}\ \emph {et~al.}(2016)\citenamefont
  {Hindmarsh}, \citenamefont {Rummukainen},\ and\ \citenamefont
  {Weir}}]{Hindmarsh:2016lhy}%
  \BibitemOpen
  \bibfield  {author} {\bibinfo {author} {\bibfnamefont {M.}~\bibnamefont
  {Hindmarsh}}, \bibinfo {author} {\bibfnamefont {K.}~\bibnamefont
  {Rummukainen}}, \ and\ \bibinfo {author} {\bibfnamefont {D.~J.}\ \bibnamefont
  {Weir}},\ }\href {\doibase 10.1103/PhysRevLett.117.251601} {\bibfield
  {journal} {\bibinfo  {journal} {Phys. Rev. Lett.}\ }\textbf {\bibinfo
  {volume} {117}},\ \bibinfo {pages} {251601} (\bibinfo {year} {2016})},\
  \Eprint {http://arxiv.org/abs/1607.00764} {arXiv:1607.00764 [hep-th]}
  \BibitemShut {NoStop}%
%%CITATION = ARXIV:1607.00764;%%
\bibitem [{\citenamefont {Blanco-Pillado}\ and\ \citenamefont
  {Olum}(2010)}]{BlancoPillado:2007zr}%
  \BibitemOpen
  \bibfield  {author} {\bibinfo {author} {\bibfnamefont {J.~J.}\ \bibnamefont
  {Blanco-Pillado}}\ and\ \bibinfo {author} {\bibfnamefont {K.~D.}\
  \bibnamefont {Olum}},\ }\href {\doibase 10.1088/1475-7516/2010/05/014}
  {\bibfield  {journal} {\bibinfo  {journal} {JCAP}\ }\textbf {\bibinfo
  {volume} {1005}},\ \bibinfo {pages} {014} (\bibinfo {year} {2010})},\ \Eprint
  {http://arxiv.org/abs/0707.3460} {arXiv:0707.3460 [astro-ph]} \BibitemShut
  {NoStop}%
%%CITATION = ARXIV:0707.3460;%%
\bibitem [{\citenamefont {Forgacs}\ \emph {et~al.}(2005)\citenamefont
  {Forgacs}, \citenamefont {Obadia},\ and\ \citenamefont
  {Reuillon}}]{Forgacs:2005vx}%
  \BibitemOpen
  \bibfield  {author} {\bibinfo {author} {\bibfnamefont {P.}~\bibnamefont
  {Forgacs}}, \bibinfo {author} {\bibfnamefont {N.}~\bibnamefont {Obadia}}, \
  and\ \bibinfo {author} {\bibfnamefont {S.}~\bibnamefont {Reuillon}},\ }\href
  {\doibase 10.1103/PhysRevD.71.035002, 10.1103/PhysRevD.71.119902} {\bibfield
  {journal} {\bibinfo  {journal} {Phys.Rev.}\ }\textbf {\bibinfo {volume}
  {D71}},\ \bibinfo {pages} {035002} (\bibinfo {year} {2005})},\ \Eprint
  {http://arxiv.org/abs/hep-th/0412057} {arXiv:hep-th/0412057 [hep-th]}
  \BibitemShut {NoStop}%
%%CITATION = HEP-TH/0412057;%%
\bibitem [{\citenamefont {Daverio}\ \emph {et~al.}(2016)\citenamefont
  {Daverio}, \citenamefont {Hindmarsh}, \citenamefont {Kunz}, \citenamefont
  {Lizarraga},\ and\ \citenamefont {Urrestilla}}]{Daverio:2015nva}%
  \BibitemOpen
  \bibfield  {author} {\bibinfo {author} {\bibfnamefont {D.}~\bibnamefont
  {Daverio}}, \bibinfo {author} {\bibfnamefont {M.}~\bibnamefont {Hindmarsh}},
  \bibinfo {author} {\bibfnamefont {M.}~\bibnamefont {Kunz}}, \bibinfo {author}
  {\bibfnamefont {J.}~\bibnamefont {Lizarraga}}, \ and\ \bibinfo {author}
  {\bibfnamefont {J.}~\bibnamefont {Urrestilla}},\ }\href {\doibase
  10.1103/PhysRevD.93.085014} {\bibfield  {journal} {\bibinfo  {journal} {Phys.
  Rev.}\ }\textbf {\bibinfo {volume} {D93}},\ \bibinfo {pages} {085014}
  (\bibinfo {year} {2016})},\ \Eprint {http://arxiv.org/abs/1510.05006}
  {arXiv:1510.05006 [astro-ph.CO]} \BibitemShut {NoStop}%
%%CITATION = ARXIV:1510.05006;%%
\bibitem [{\citenamefont {Rajantie}(2009)}]{Rajantie:2008bc}%
  \BibitemOpen
  \bibfield  {author} {\bibinfo {author} {\bibfnamefont {A.}~\bibnamefont
  {Rajantie}},\ }\href {\doibase 10.1103/PhysRevD.79.043515} {\bibfield
  {journal} {\bibinfo  {journal} {Phys. Rev.}\ }\textbf {\bibinfo {volume}
  {D79}},\ \bibinfo {pages} {043515} (\bibinfo {year} {2009})},\ \Eprint
  {http://arxiv.org/abs/0810.3007} {arXiv:0810.3007 [astro-ph]} \BibitemShut
  {NoStop}%
%%CITATION = ARXIV:0810.3007;%%
\bibitem [{\citenamefont {Arnold}(1997)}]{Arnold:1997yb}%
  \BibitemOpen
  \bibfield  {author} {\bibinfo {author} {\bibfnamefont {P.~B.}\ \bibnamefont
  {Arnold}},\ }\href {\doibase 10.1103/PhysRevD.55.7781} {\bibfield  {journal}
  {\bibinfo  {journal} {Phys. Rev.}\ }\textbf {\bibinfo {volume} {D55}},\
  \bibinfo {pages} {7781} (\bibinfo {year} {1997})},\ \Eprint
  {http://arxiv.org/abs/hep-ph/9701393} {arXiv:hep-ph/9701393 [hep-ph]}
  \BibitemShut {NoStop}%
%%CITATION = HEP-PH/9701393;%%
\bibitem [{\citenamefont {Moore}(1996)}]{Moore:1996wn}%
  \BibitemOpen
  \bibfield  {author} {\bibinfo {author} {\bibfnamefont {G.~D.}\ \bibnamefont
  {Moore}},\ }\href {\doibase 10.1016/S0550-3213(96)00497-X} {\bibfield
  {journal} {\bibinfo  {journal} {Nucl. Phys.}\ }\textbf {\bibinfo {volume}
  {B480}},\ \bibinfo {pages} {689} (\bibinfo {year} {1996})},\ \Eprint
  {http://arxiv.org/abs/hep-lat/9605001} {arXiv:hep-lat/9605001 [hep-lat]}
  \BibitemShut {NoStop}%
%%CITATION = HEP-LAT/9605001;%%
\bibitem [{\citenamefont {Hindmarsh}\ \emph {et~al.}(2014)\citenamefont
  {Hindmarsh}, \citenamefont {Rummukainen}, \citenamefont {Tenkanen},\ and\
  \citenamefont {Weir}}]{Hindmarsh:2014rka}%
  \BibitemOpen
  \bibfield  {author} {\bibinfo {author} {\bibfnamefont {M.}~\bibnamefont
  {Hindmarsh}}, \bibinfo {author} {\bibfnamefont {K.}~\bibnamefont
  {Rummukainen}}, \bibinfo {author} {\bibfnamefont {T.~V.~I.}\ \bibnamefont
  {Tenkanen}}, \ and\ \bibinfo {author} {\bibfnamefont {D.~J.}\ \bibnamefont
  {Weir}},\ }\href {\doibase 10.1103/PhysRevD.90.043539} {\bibfield  {journal}
  {\bibinfo  {journal} {Phys. Rev.}\ }\textbf {\bibinfo {volume} {D90}},\
  \bibinfo {pages} {043539} (\bibinfo {year} {2014})},\ \Eprint
  {http://arxiv.org/abs/1406.1688} {arXiv:1406.1688 [hep-lat]} \BibitemShut
  {NoStop}%
%%CITATION = ARXIV:1406.1688;%%
\bibitem [{\citenamefont {Davis}\ \emph {et~al.}(2000)\citenamefont {Davis},
  \citenamefont {Kibble}, \citenamefont {Rajantie},\ and\ \citenamefont
  {Shanahan}}]{Davis:2000kv}%
  \BibitemOpen
  \bibfield  {author} {\bibinfo {author} {\bibfnamefont {A.}~\bibnamefont
  {Davis}}, \bibinfo {author} {\bibfnamefont {T.}~\bibnamefont {Kibble}},
  \bibinfo {author} {\bibfnamefont {A.}~\bibnamefont {Rajantie}}, \ and\
  \bibinfo {author} {\bibfnamefont {H.}~\bibnamefont {Shanahan}},\ }\href@noop
  {} {\bibfield  {journal} {\bibinfo  {journal} {JHEP}\ }\textbf {\bibinfo
  {volume} {0011}},\ \bibinfo {pages} {010} (\bibinfo {year} {2000})},\ \Eprint
  {http://arxiv.org/abs/hep-lat/0009037} {arXiv:hep-lat/0009037 [hep-lat]}
  \BibitemShut {NoStop}%
%%CITATION = HEP-LAT/0009037;%%
\bibitem [{\citenamefont {Scherrer}\ and\ \citenamefont
  {Vilenkin}(1998)}]{Scherrer:1997sq}%
  \BibitemOpen
  \bibfield  {author} {\bibinfo {author} {\bibfnamefont {R.~J.}\ \bibnamefont
  {Scherrer}}\ and\ \bibinfo {author} {\bibfnamefont {A.}~\bibnamefont
  {Vilenkin}},\ }\href {\doibase 10.1103/PhysRevD.58.103501} {\bibfield
  {journal} {\bibinfo  {journal} {Phys. Rev.}\ }\textbf {\bibinfo {volume}
  {D58}},\ \bibinfo {pages} {103501} (\bibinfo {year} {1998})},\ \Eprint
  {http://arxiv.org/abs/hep-ph/9709498} {arXiv:hep-ph/9709498 [hep-ph]}
  \BibitemShut {NoStop}%
%%CITATION = HEP-PH/9709498;%%
\bibitem [{\citenamefont {Ostriker}\ \emph {et~al.}(1986)\citenamefont
  {Ostriker}, \citenamefont {Thompson},\ and\ \citenamefont
  {Witten}}]{Ostriker:1986xc}%
  \BibitemOpen
  \bibfield  {author} {\bibinfo {author} {\bibfnamefont {J.~P.}\ \bibnamefont
  {Ostriker}}, \bibinfo {author} {\bibfnamefont {A.~C.}\ \bibnamefont
  {Thompson}}, \ and\ \bibinfo {author} {\bibfnamefont {E.}~\bibnamefont
  {Witten}},\ }\href {\doibase 10.1016/0370-2693(86)90301-1} {\bibfield
  {journal} {\bibinfo  {journal} {Phys. Lett.}\ }\textbf {\bibinfo {volume}
  {B180}},\ \bibinfo {pages} {231} (\bibinfo {year} {1986})}\BibitemShut
  {NoStop}%
%%CITATION = PHLTA,B180,231;%%
\bibitem [{\citenamefont {Copeland}\ \emph {et~al.}(1987)\citenamefont
  {Copeland}, \citenamefont {Turok},\ and\ \citenamefont
  {Hindmarsh}}]{Copeland:1987th}%
  \BibitemOpen
  \bibfield  {author} {\bibinfo {author} {\bibfnamefont {E.~J.}\ \bibnamefont
  {Copeland}}, \bibinfo {author} {\bibfnamefont {N.}~\bibnamefont {Turok}}, \
  and\ \bibinfo {author} {\bibfnamefont {M.}~\bibnamefont {Hindmarsh}},\ }\href
  {\doibase 10.1103/PhysRevLett.58.1910} {\bibfield  {journal} {\bibinfo
  {journal} {Phys. Rev. Lett.}\ }\textbf {\bibinfo {volume} {58}},\ \bibinfo
  {pages} {1910} (\bibinfo {year} {1987})}\BibitemShut {NoStop}%
%%CITATION = PRLTA,58,1910;%%
\bibitem [{\citenamefont {Davis}\ and\ \citenamefont
  {Shellard}(1989)}]{Davis:1988ij}%
  \BibitemOpen
  \bibfield  {author} {\bibinfo {author} {\bibfnamefont {R.~L.}\ \bibnamefont
  {Davis}}\ and\ \bibinfo {author} {\bibfnamefont {E.~P.~S.}\ \bibnamefont
  {Shellard}},\ }\href {\doibase 10.1016/0550-3213(89)90594-4} {\bibfield
  {journal} {\bibinfo  {journal} {Nucl. Phys.}\ }\textbf {\bibinfo {volume}
  {B323}},\ \bibinfo {pages} {209} (\bibinfo {year} {1989})}\BibitemShut
  {NoStop}%
%%CITATION = NUPHA,B323,209;%%
\bibitem [{\citenamefont {Santana~Mota}\ and\ \citenamefont
  {Hindmarsh}(2015)}]{Mota:2014uka}%
  \BibitemOpen
  \bibfield  {author} {\bibinfo {author} {\bibfnamefont {H.~F.}\ \bibnamefont
  {Santana~Mota}}\ and\ \bibinfo {author} {\bibfnamefont {M.}~\bibnamefont
  {Hindmarsh}},\ }\href {\doibase 10.1103/PhysRevD.91.043001} {\bibfield
  {journal} {\bibinfo  {journal} {Phys. Rev.}\ }\textbf {\bibinfo {volume}
  {D91}},\ \bibinfo {pages} {043001} (\bibinfo {year} {2015})},\ \Eprint
  {http://arxiv.org/abs/1407.3599} {arXiv:1407.3599 [hep-ph]} \BibitemShut
  {NoStop}%
%%CITATION = ARXIV:1407.3599;%%
\bibitem [{\citenamefont {Moss}\ and\ \citenamefont
  {Pogosian}(2014)}]{Moss:2014cra}%
  \BibitemOpen
  \bibfield  {author} {\bibinfo {author} {\bibfnamefont {A.}~\bibnamefont
  {Moss}}\ and\ \bibinfo {author} {\bibfnamefont {L.}~\bibnamefont
  {Pogosian}},\ }\href {\doibase 10.1103/PhysRevLett.112.171302} {\bibfield
  {journal} {\bibinfo  {journal} {Phys. Rev. Lett.}\ }\textbf {\bibinfo
  {volume} {112}},\ \bibinfo {pages} {171302} (\bibinfo {year} {2014})},\
  \Eprint {http://arxiv.org/abs/1403.6105} {arXiv:1403.6105 [astro-ph.CO]}
  \BibitemShut {NoStop}%
%%CITATION = ARXIV:1403.6105;%%
\bibitem [{\citenamefont {Charnock}\ \emph {et~al.}(2016)\citenamefont
  {Charnock}, \citenamefont {Avgoustidis}, \citenamefont {Copeland},\ and\
  \citenamefont {Moss}}]{Charnock:2016nzm}%
  \BibitemOpen
  \bibfield  {author} {\bibinfo {author} {\bibfnamefont {T.}~\bibnamefont
  {Charnock}}, \bibinfo {author} {\bibfnamefont {A.}~\bibnamefont
  {Avgoustidis}}, \bibinfo {author} {\bibfnamefont {E.~J.}\ \bibnamefont
  {Copeland}}, \ and\ \bibinfo {author} {\bibfnamefont {A.}~\bibnamefont
  {Moss}},\ }\href {\doibase 10.1103/PhysRevD.93.123503} {\bibfield  {journal}
  {\bibinfo  {journal} {Phys. Rev.}\ }\textbf {\bibinfo {volume} {D93}},\
  \bibinfo {pages} {123503} (\bibinfo {year} {2016})},\ \Eprint
  {http://arxiv.org/abs/1603.01275} {arXiv:1603.01275 [astro-ph.CO]}
  \BibitemShut {NoStop}%
%%CITATION = ARXIV:1603.01275;%%
\bibitem [{\citenamefont {Lizarraga}\ \emph {et~al.}(2016)\citenamefont
  {Lizarraga}, \citenamefont {Urrestilla}, \citenamefont {Daverio},
  \citenamefont {Hindmarsh},\ and\ \citenamefont {Kunz}}]{Lizarraga:2016onn}%
  \BibitemOpen
  \bibfield  {author} {\bibinfo {author} {\bibfnamefont {J.}~\bibnamefont
  {Lizarraga}}, \bibinfo {author} {\bibfnamefont {J.}~\bibnamefont
  {Urrestilla}}, \bibinfo {author} {\bibfnamefont {D.}~\bibnamefont {Daverio}},
  \bibinfo {author} {\bibfnamefont {M.}~\bibnamefont {Hindmarsh}}, \ and\
  \bibinfo {author} {\bibfnamefont {M.}~\bibnamefont {Kunz}},\ }\href {\doibase
  10.1088/1475-7516/2016/10/042} {\bibfield  {journal} {\bibinfo  {journal}
  {JCAP}\ }\textbf {\bibinfo {volume} {1610}},\ \bibinfo {pages} {042}
  (\bibinfo {year} {2016})},\ \Eprint {http://arxiv.org/abs/1609.03386}
  {arXiv:1609.03386 [astro-ph.CO]} \BibitemShut {NoStop}%
%%CITATION = ARXIV:1609.03386;%%
\bibitem [{\citenamefont {Siemens}\ \emph {et~al.}(2001)\citenamefont
  {Siemens}, \citenamefont {Martin},\ and\ \citenamefont
  {Olum}}]{Siemens:2000ty}%
  \BibitemOpen
  \bibfield  {author} {\bibinfo {author} {\bibfnamefont {X.}~\bibnamefont
  {Siemens}}, \bibinfo {author} {\bibfnamefont {X.}~\bibnamefont {Martin}}, \
  and\ \bibinfo {author} {\bibfnamefont {K.~D.}\ \bibnamefont {Olum}},\ }\href
  {\doibase 10.1016/S0550-3213(00)00672-6} {\bibfield  {journal} {\bibinfo
  {journal} {Nucl.Phys.}\ }\textbf {\bibinfo {volume} {B595}},\ \bibinfo
  {pages} {402} (\bibinfo {year} {2001})},\ \Eprint
  {http://arxiv.org/abs/astro-ph/0005411} {arXiv:astro-ph/0005411 [astro-ph]}
  \BibitemShut {NoStop}%
%%CITATION = ASTRO-PH/0005411;%%
\bibitem [{\citenamefont {Lentati}\ \emph {et~al.}(2015)\citenamefont {Lentati}
  \emph {et~al.}}]{Lentati:2015qwp}%
  \BibitemOpen
  \bibfield  {author} {\bibinfo {author} {\bibfnamefont {L.}~\bibnamefont
  {Lentati}} \emph {et~al.},\ }\href {\doibase 10.1093/mnras/stv1538}
  {\bibfield  {journal} {\bibinfo  {journal} {Mon. Not. Roy. Astron. Soc.}\
  }\textbf {\bibinfo {volume} {453}},\ \bibinfo {pages} {2576} (\bibinfo {year}
  {2015})},\ \Eprint {http://arxiv.org/abs/1504.03692} {arXiv:1504.03692
  [astro-ph.CO]} \BibitemShut {NoStop}%
%%CITATION = ARXIV:1504.03692;%%
\bibitem [{\citenamefont {Arzoumanian}\ \emph {et~al.}(2016)\citenamefont
  {Arzoumanian} \emph {et~al.}}]{Arzoumanian:2015liz}%
  \BibitemOpen
  \bibfield  {author} {\bibinfo {author} {\bibfnamefont {Z.}~\bibnamefont
  {Arzoumanian}} \emph {et~al.} (\bibinfo {collaboration} {NANOGrav}),\ }\href
  {\doibase 10.3847/0004-637X/821/1/13} {\bibfield  {journal} {\bibinfo
  {journal} {Astrophys. J.}\ }\textbf {\bibinfo {volume} {821}},\ \bibinfo
  {pages} {13} (\bibinfo {year} {2016})},\ \Eprint
  {http://arxiv.org/abs/1508.03024} {arXiv:1508.03024 [astro-ph.GA]}
  \BibitemShut {NoStop}%
%%CITATION = ARXIV:1508.03024;%%
\bibitem [{\citenamefont {Crank}\ and\ \citenamefont
  {Nicolson}(1996)}]{Crank1996}%
  \BibitemOpen
  \bibfield  {author} {\bibinfo {author} {\bibfnamefont {J.}~\bibnamefont
  {Crank}}\ and\ \bibinfo {author} {\bibfnamefont {P.}~\bibnamefont
  {Nicolson}},\ }\href {\doibase 10.1007/BF02127704} {\bibfield  {journal}
  {\bibinfo  {journal} {Advances in Computational Mathematics}\ }\textbf
  {\bibinfo {volume} {6}},\ \bibinfo {pages} {207} (\bibinfo {year}
  {1996})}\BibitemShut {NoStop}%
\bibitem [{\citenamefont {Edwards}\ \emph {et~al.}(2009)\citenamefont
  {Edwards}, \citenamefont {Mehta}, \citenamefont {Rajantie},\ and\
  \citenamefont {von Smekal}}]{Edwards:2009bw}%
  \BibitemOpen
  \bibfield  {author} {\bibinfo {author} {\bibfnamefont {S.}~\bibnamefont
  {Edwards}}, \bibinfo {author} {\bibfnamefont {D.~B.}\ \bibnamefont {Mehta}},
  \bibinfo {author} {\bibfnamefont {A.}~\bibnamefont {Rajantie}}, \ and\
  \bibinfo {author} {\bibfnamefont {L.}~\bibnamefont {von Smekal}},\ }\href
  {\doibase 10.1103/PhysRevD.80.065030} {\bibfield  {journal} {\bibinfo
  {journal} {Phys.Rev.}\ }\textbf {\bibinfo {volume} {D80}},\ \bibinfo {pages}
  {065030} (\bibinfo {year} {2009})},\ \Eprint {http://arxiv.org/abs/0906.5531}
  {arXiv:0906.5531 [hep-lat]} \BibitemShut {NoStop}%
%%CITATION = ARXIV:0906.5531;%%
\end{thebibliography}%

\end{document}